\documentstyle[12pt,fleqn]{article}

\def\thebibliography#1{\bigskip\section*{\centering
References\\}\bigskip\list
  {\arabic{enumi}.}{\settowidth\labelwidth{#1}\leftmargin\labelwidth
    \advance\leftmargin\labelsep
    \usecounter{enumi}}
    \def\newblock{\hskip .11em plus .33em minus .07em}
    \sloppy\clubpenalty4000\widowpenalty4000
    \sfcode`\.=1000\relax}

\let\Large=\large

\newcommand{\bx}[1]{{\rm\fbox{$ #1$}}}

\def\op#1{\mathop{\fam0 #1}\limits}

\newcommand{\pr}{{\rm pr}}

\newcommand{\nm}[1]{\mid{#1}\mid}

\newcommand{\beq}{\begin{equation}}
\newcommand{\eeq}{\end{equation}}
\newcommand{\ben}{\begin{eqnarray}}
\newcommand{\een}{\end{eqnarray}}
\newcommand{\be}{\begin{eqnarray*}}
\newcommand{\ee}{\end{eqnarray*}}
\newcommand{\bea}{\begin{eqalph}}
\newcommand{\eea}{\end{eqalph}}

\newcommand{\cL}{{\cal L}}

\newcommand{\cD}{{\cal D}}
\newcommand{\bL}{{\bf L}}
\newcommand{\rL}{{\rm L}}
\newcommand{\rs}{{\rm s}}
\newcommand{\al}{\alpha}

\newcommand{\bt}{\beta}
\newcommand{\dl}{\delta}
\newcommand{\la}{\lambda}

\newcommand{\f}{\phi}

\newcommand{\F}{\Phi}

\newcommand{\om}{\omega}

\newcommand{\m}{\mu}
\newcommand{\n}{\nu}
\newcommand{\g}{\gamma}

\newcommand{\e}{\epsilon}

\newcommand{\th}{\theta}

\newcommand{\si}{\sigma}
\newcommand{\Si}{\Sigma}

\newcommand{\w}{\wedge}
\newcommand{\wt}{\widetilde}
\newcommand{\wh}{\widehat}
\newcommand{\ol}{\overline}
\newcommand{\dr}{\partial}

\newcommand{\ar}{\op\longrightarrow}

\newcommand{\ot}{\otimes}

\let\ssection=\section
\renewcommand{\section}{\setcounter{equation}{0}\ssection}

\newcounter{eqalph}[section]
\newcounter{equationa}[section]
\newcounter{example}
\newcounter{remark}
\newcounter{theorem}
\newcounter{proposition}
\newcounter{definition}

\def\theremark{\arabic{remark}}

\def\thedefinition{\arabic{definition}}

\newenvironment{proof}{\noindent {\bf Proof.}}{{\Large $\bullet$} \bigskip}
\newenvironment{ex}{\refstepcounter{remark} \medskip\noindent{\bf Example
\theremark.}}{{\Large $\bullet$} \bigskip }
\newenvironment{rem}{\refstepcounter{remark} \medskip\noindent{\bf Remark
\theremark.}}{{\Large $\bullet$} \bigskip }
\newenvironment{theo}{\refstepcounter{definition} \bigskip\noindent{\sc
Theorem \thedefinition}.}{$\Box$\bigskip }
\newenvironment{prop}{\refstepcounter{definition} \bigskip\noindent{\sc
Proposition \thedefinition}.}{$\Box$ \bigskip }

\newenvironment{eqalph}{\stepcounter{equation}
\setcounter{equationa}{\value{equation}}
\setcounter{equation}{0}

\begin{eqnarray}}{\end{eqnarray}
\setcounter{equation}{\value{equationa}}}

\begin{document}
\hbox{}

\begin{center}

{\Large \bf Universal Spin Structure 
\medskip

in Gauge Gravitation Theory}
\bigskip

{\sc G. Giachetta, L. Mangiarotti,}
\medskip

Department of Mathematics and Physics, University of Camerino,

62032 Camerino (MC), Italy 

E-mail: mangiaro@camserv.unicam.it
\medskip

{\sc G. Sardanashvily}
\medskip

Department of Theoretical Physics, Moscow State University

117234 Moscow, Russia

E-mail: sard@grav.phys.msu.su
\end{center}

\bigskip

\begin{abstract}
Building on the universal covering group of the general linear group, we
introduce the composite spinor bundle whose subbundles are Lorentz spin
structures associated with different gravitational fields. General covariant
transformations of this composite spinor bundle are canonically defined.
\end{abstract}

\section{Introduction}

Einstein's General Relativity and metric-affine gravitation theory 
are formulated on fibre bundles
$Y\to X$ which admit a canonical
lift of every diffeomorphism of the  base $X$. These are called bundles of
geometric objects. The lift obtained is a general covariant transformation of
$Y$. The invariance of a gravitational Lagrangian density under these
transformations leads to the energy-momentum conservation laws where the
gravitational energy-momentum flow is reduced to the generalized Komar
superpotential
\cite{bor,giach95,giachcqg,giach97,nov,nov93,sard97b}. 

Difficulties arise in the gauge gravitation theory because of Dirac's fermion
fields. The corresponding Lorentz spin structure is associated with a certain
gravitational field, and it is not preserved under 
general covariant transformations. To overcome this difficulty, the
universal spin structure on a world manifold is here introduced.

Consider the universal two-fold covering group $\wt{GL_4}$ of the general
linear group
\be
GL_4={\rm GL}^+(4,{\bf R})
\ee
and the corresponding two-fold covering bundle $\wt{LX}$ of the bundle of
linear frames $LX$ \cite{dabr,perc,swit}. There is the commutative diagram
\beq
\begin{array}{ccc}
 \wt{GL}_4 & \longrightarrow &  GL_4 \\
 \put(0,-10){\vector(0,1){20}} & 
& \put(0,-10){\vector(0,1){20}}  \\
L_\rs & \ar & {\rm L} 
\end{array} 
\eeq
where $L=SO^0(1,3)$ is the proper Lorentz group and $L_s=SL(2,{\bf C})$ its
two-fold covering spin group.  One can consider the spinor representations
of the group $\wt{GL}_4$ which, however, are infinite-dimensional
\cite{heh,nee}. 

Here we pursue a different approach as follows.
The total space of the $\wt{GL_4}$-principal bundle 
$\wt{LX}\to X$ is the $L_s$-principal bundle $\wt{LX}\to \Si_{\rm T}$ 
over the
quotient bundle
\be
\Si_{\rm T}:=\wt{LX}/L_s \to X
\ee
whose sections are tetrad gravitational fields $h$.

Let us consider the Lorentz spinor bundle 
\be
S=(\wt{LX}\times V)/L_s
\ee
associated with the principal bundle $\wt{LX}\to \Si_{\rm T}$. Given a
tetrad field $h$, the restriction of this bundle to $h(X)\subset \Si_{\rm T}$
is a subbundle of the composite spinor bundle
\be
S\to \Si_{\rm T} \to X
\ee
which is exactly the Lorentz spin structure associated with the
gravitational field $h$. General covariant transformations 
of the linear frame
bundle $LX$ and, consequently, of the bundles $\wt{LX}$ 
and $S$ are canonically
defined.

\section{Preliminairies}

Throughout the paper manifolds are real, finite-dimensional,
Hausdorff, second-countable (hence paracompact) and connected. By a world
manifold
$X$ is meant a 4-dimensional manifold, which is assumed to be non-compact,
orientable and parallelizable in order that a
pseudo-Riemannian metric, a spin structure and a causal space-time structure
can exist on it
\cite{ger,wist}. An orientation on $X$ is chosen.

\begin{rem}
In classical field theory, if cosmological models are not discussed, 
some causality conditions should be satisfied (see \cite{haw}). A compact
space-time does not possess this property because it has closed time-like
curves. For these reasons, we restrict our considerations to non-compact
manifolds. Every non-compact manifold admits a 
non-zero vector field and, as a
consequence, a pseudo-Riemannian metric
(\cite{dods}, p.167). A non-compact manifold $X$ has a spin structure iff it
is parallelizable (i.e., the tangent bundle $TX\to X$ is trivial) \cite{ger}.
Moreover, the spin structure on a non-compact 
parallelizable manifold is unique
\cite{avis,ger}. The orientability of a world manifold is not needed 
for a pseudo-Riemannian structure and a spin structure to exist. This
requirement and the additional condition of 
time-orientability seem natural if
we are not concerned with cosmological models \cite{dods}. Note that the
requirement of a manifold to be paracompact also has a physical meaning. A
manifold is paracompact iff it admits a Riemannian structure (\cite{marath};
\cite{kob}, p.271).
\end{rem}

A linear
connection $K$ and a pseudo-Riemannian metric
$g$ on a world manifold $X$ are said to be
a world connection and a world metric,
respectively. We recall the coordinate expressions of a world connection on
$TX$ and $T^*X$, respectively,
\ben
&& K= dx^\la\otimes (\dr_\la +K_\la{}^\m{}_\n
\dot x^\n
\frac{\dr}{\dr\dot x^\m}), \label{08} \\
&& K^*= dx^\la\otimes (\dr_\la -K_\la{}^\m{}_\n \dot x_\m
\frac{\dr}{\dr\dot x_\n}). \label{08'}
\een

\begin{rem}
Unless otherwise stated, the coordinate atlas $\Psi_X=\{(U_\zeta,\f_\zeta)\}$
of $X$, the corresponding  holonomic atlas 
\beq
\Psi_T= \{(U_\zeta,T\f_\zeta)\} \label{b3241}
\eeq
of the tangent bundle $TX$ and the holonomic atlases of its
associated fibre bundles are assumed to be fixed.
\end{rem}

Let 
\be
\pi_{LX}:LX\to X
\ee
be the principal bundle of oriented linear
frames in the tangent spaces to a world manifold $X$ (or simply the  frame
bundle). Its structure group is 
\be
GL_4={\rm GL}^+(4,{\bf R}).
\ee

By definition, a world manifold $X$ is parallelizable if the
frame bundle
$LX\to X$ is trivial and, consequently, it admits global sections (i.e., 
global frame fields). 

Considering the holonomic frames
$\{\dr_\m\}$ in the tangent bundle $TX$ associated with the holonomic atlas
(\ref{b3241}), every element
$\{H_a\}$ of the frame bundle
$LX$ takes the form 
\be
H_a=H^\m{}_a\dr_\m,
\ee
where the matrix $H^\m{}_a$ belongs to the group $GL_4$. 
The frame bundle $LX$
is provided with the bundle coordinates 
\be
(x^\la, H^\m{}_a), \qquad H'^\m{}_a=\frac{\dr x'^\m}{\dr x^\la}H^\la{}_a. 
\ee
In these coordinates, the canonical action  
of the structure group $GL_4$ on $LX$ on the right reads 
\be
R_M: H^\m{}_a\mapsto H^\m{}_bM^b{}_a, \qquad M\in GL_4.
\ee

The frame bundle $LX$ is equipped with the canonical ${\bf
R}^4$-valued 1-form $\th_{LX}$. Its coordinate
expression is
\beq
\th_{LX} = H^b{}_\m dx^\m\ot t_a,\label{b3133'}
\eeq
where $\{t_a\}$ is a fixed basis of ${\bf R}^4$ and $H^b{}_\m$ is the inverse
matrix of $H^\m{}_a$.

The important peculiarity of the frame bundle $LX\to X$ is that
every diffeomorphism $f$ of $X$ gives rise canonically to the automorphism
\beq
\wt f: (x^\la, H^\la{}_a)\mapsto (f^\la(x),\dr_\m f^\la H^\m{}_a) \label{025}
\eeq
of $LX$ and to the corresponding
automorphisms (denoted by the same symbol)
\be
\wt f:T=(LX\times V)/GL_4\to (\wt f(LX)\times V)/GL_4
\ee
of any fibre bundle $T$ canonically associated with $LX$.
These automorphisms are called general covariant
transformations.

\begin{ex} If $T=TX$, the lift $\wt f=Tf$ 
is the familiar tangent morphism to the diffeomorphism $f$.
If $T=T^*X$ is the cotangent bundle, $\wt f=T^*f^{-1}$ is the dual of the
linear morphism $Tf^{-1}$.
\end{ex}

The lift (\ref{025}) provides the  canonical horizontal lift
 $\wt\tau$ of every vector
field $\tau$ on
$X$ over the principal bundle $LX$ and over the associated fibre bundles. The
canonical lift of $\tau$ over $LX$ is defined by the relation
\be
\bL_{\wt\tau}\th_{LX}=0.
\ee
We have the corresponding
canonical lift 
\beq
\wt\tau = \tau^\m\dr_\m + [\dr_\nu\tau^{\al_1}\dot
x^{\nu\al_2\cdots\al_m}_{\bt_1\cdots\bt_k} + \ldots
-\dr_{\bt_1}\tau^\nu \dot x^{\al_1\cdots\al_m}_{\nu\bt_2\cdots\bt_k}
-\ldots]\frac{\dr}{\dr \dot
x^{\al_1\cdots\al_m}_{\bt_1\cdots\bt_k}} \label{l28}
\eeq
of $\tau$ over the tensor bundle 
\beq
T=(\op\ot^mTX)\ot(\op\ot^kT^*X). \label{971}
\eeq
and, in particular, the lifts
\beq
\wt\tau = \tau^\m\dr_\m +\dr_\nu\tau^\al\dot x^\nu\frac{\dr}{\dr\dot x^\al}
\label{l27}
\eeq
over the tangent bundle $TX$ and
\beq 
\wt\tau = \tau^\m\dr_\m -\dr_\bt\tau^\nu\dot x_\nu\frac{\dr}{\dr\dot x_\bt}
\label{l27'} 
\eeq
over the cotangent bundle $T^*X$. 

\begin{ex}
A  pseudo-Riemannian metric $g$
 on a world manifold
$X$ is represented by a section of the fibre bundle 
\beq
\Si_{\rm PR}= GLX/{\rm O}(1,3),\label{b3203}
\eeq
where  by $GLX$ is meant the bundle of all
linear frames in $TX$ and O$(1,3)$ is the complete Lorentz group. 
We call $\Si_{\rm PR}$ the  metric bundle.
 Since $X$ is oriented, $\Si_{\rm PR}$ is
associated with the principal bundle $LX$ of oriented frames in $TX$.
Its typical fibre
is the quotient
\be
{\rm GL}(4,{\bf R})/{\rm O}(1,3).
\ee
This
quotient space is homotopic to the Grassman manifold 
${\bf G}(4,3;{\bf R})$
and is homeomorphic to the topological space
${\bf RP}^3\times
{\bf R}^7$ (\cite{perc}, p.164), where by ${\bf RP}^3$ is meant the
3-dimensional real projective space. 
 
For the sake of siplicity, we identify the metric
bundle with an open subbundle of the tensor bundle
\be
\Si_{\rm PR}\subset \op\vee^2 TX.
\ee
This is coordinatized by $(x^\la, \si^{\m\nu})$. 
As usual, by $\si_{\m\n}$ are
meant the components of the inverse matrix, and  $\si =\det(\si_{\m\n})$.

The canonical lift $\wt\tau$ onto $\Si_{\rm PR}$ of a vector field $\tau$ on
$X$ reads
\beq
\wt\tau =\tau^\la\dr_\la + 
(\dr_\nu\tau^\al\si^{\nu\bt}
+\dr_\nu\tau^\bt\si^{\nu\al})\frac{\dr}{\dr\si^{\al\bt}}.
\label{973}
\eeq
\end{ex}

\begin{ex}
Since the world connections are induced by the principal connections on
the linear frame bundle $LX$, there is the one-to-one correspondence
between the world connections and the sections of 
the quotient fibre bundle 
\beq
C_K=J^1LX/GL_4, \label{015}
\eeq
where by $J^1LX$ is meant the first order jet manifold of the fibre bundle
$LX\to X$. We call (\ref{015}) the  bundle of world connections.

With respect to the holonomic frames in $TX$, the bundle $C_K$ 
is coordinatized by $(x^\la, k_\la{}^\nu{}_{\al})$ 
so that, for any section $K$
of $C_K\to X$,
\be
k_\la{}^\nu{}_\al\circ K=K_\la{}^\nu{}_\al
\ee
are the coefficients of the world connection $K$ (\ref{08}). 

The bundle $C_K$ fails to be associated with $LX$, 
but it is an affine bundle
modelled on a vector bundle associated with
$LX$. There exists the canonical lift
\beq
\wt\tau = \tau^\m\dr_\m +[\dr_\nu\tau^\al k_\m{}^\nu{}_\bt -
\dr_\bt\tau^\nu k_\m{}^\al{}_\nu - \dr_\m\tau^\nu
k_\nu{}^\al{}_\bt + \dr_{\m\bt}\tau^\al]\frac{\dr}{\dr k_\m{}^\al{}_\bt}
\label{b3150}
\eeq
of a vector field $\tau$ on $X$ over $C_K$.
\end{ex}

In fact, the canonical lift $\wt\tau$ (\ref{l28}) is the 
horizontal lift of $\tau$ by
means of a symmetric world connection (\ref{08}), for which $\tau$ is 
a geodesic vector field, i.e.,
\be
\dr_\nu\tau^\al =K_\nu{}^\al{}_\bt\tau^\bt.
\ee

\begin{rem}
One can construct the horizontal lift 
\beq
\tau_K =\tau^\la(\dr_\la +K_\la{}^\bt{}_\al\dot x^\al\frac{\dr}{\dr\dot
x^\bt}) \label{b3180}
\eeq
of a vector field
$\tau$ on $X$ over $TX$ (and other tensor bundles)
by means of any world connection $K$. This is the generator of a 1-parameter
group of non-holonomic automorphisms of
the frame bundle $LX$. We meet non-holonomic automorphisms in the gauge
theory of the general linear group $GL_4$ \cite{heh}. Note that the lifts
(\ref{l27}) and (\ref{b3180}) play the 
role of generators of the gauge group
of translations in the pioneer gauge 
gravitation theories (see \cite{heh76,iva}
and references therein).
\end{rem}

\section{Dirac spinors}

We describe Dirac spinors as follows \cite{cra,obu,rodr} (see
\cite{budi,law} for a general description of the
Clifford algebra techniques).

Let $M$ be the Minkowski space equipped with the Minkowski metric 
\be
\eta ={\rm diag}(1,-1,-1,-1)
\ee
written with respect to a basis $\{e^a\}$ of $M$.

Let ${\bf C}_{1,3}$ be the complex  Clifford
algebra  generated by elements of
$M$. This is the complexified quotient of
the tensor algebra $\ot M$ of $M$ by the two-sided ideal generated by the
elements
\be
e\otimes e'+e'\otimes e-2\eta(e,e')\in \ot M,\qquad e,\ e'\in M.
\ee

\begin{rem}
The complex Clifford algebra ${\bf C}_{1,3}$ is isomorphic to the real
Clifford algebra ${\bf R}_{2,3}$,  whose generating
space is ${\bf R}^5$ with the metric 
${\rm diag}(1,-1,-1,-1,1).$
The
subalgebra generated by the elements of $M\subset {\bf
R}^5$ is the real Clifford algebra 
${\bf R}_{1,3}$.
\end{rem}

A  spinor space $V$ is defined to be a
minimal left ideal of ${\bf C}_{1,3}$  on
which this algebra acts on the left. We have the representation
\ben
&& \g: M\otimes V \to V, \label{w01}\\
&&\g(e^a)=\g^a, \nonumber 
\een
of elements of the Minkowski space $M\subset{\bf C}_{1,3}$ by the Dirac
$\g$-matrices on $V$. 

\begin{rem}
The explicit form of this representation depends on the choice of the
minimal left ideal $V$ of ${\bf C}_{1,3}$. Different 
ideals lead to equivalent
representations.
\end{rem}

The spinor space $V$ is provided with 
the  spinor metric 
\beq
a(v,v') =\frac12(v^+\g^0v' +{v'}^+\g^0v),
\label{b3201}
\eeq
where $e^0\in M$ satisfies the
conditions
\be
(e^0)^+=e^0, \qquad (e^0e)^+=e^0e, \qquad \forall e\in M.
\ee

By definition, the  Clifford group $G_{1,3}$
comprises all invertible elements $l_s$ of the real Clifford
algebra ${\bf R}_{1,3}$ such that the corresponding inner automorphisms keep
the Minkowski space $M\subset {\bf R}_{1,3}$, that is,
\beq
l_sel^{-1}_s = l(e), \qquad e\in M, \label{b3200}
\eeq
where $l\in O(1,3)$ is a Lorentz transformation of $M$. 
The automorphisms (\ref{b3200}) preserve also the
representation (\ref{w01}), i.e.,
\beq
\g (lM\otimes l_sV) = l_s\g (M\otimes V). \label{b3191}
\eeq

Thereby, we have an
epimorphism of the Clifford group ${\bf G}_{1,3}$ onto the Lorentz group
O$(1,3)$. Since the action (\ref{b3200}) of the Clifford group on the
Minkowski space 
$M$ is not effective, one usually considers its  pin
and spin subgroups.  
The 
subgroup Pin$(1,3)$ of $G_{1,3}$ is generated by elements $e\in
M$ such that $\eta(e,e)=\pm 1$.  

The even part of 
Pin$(1,3)$ is the spin group ${\rm Spin}(1,3)$. 
Its component of the unity 
\be
L_\rs={\rm Spin}^0(1,3)\simeq {\rm SL}(2,{\bf C})
\ee
is the
well-known  two-fold universal covering group
\beq
z_L:L_\rs\to \rL=L_\rs/{\bf Z}_2, \qquad {\bf Z}_2=\{1,-1\}, \label{b3204}
\eeq
of the proper  
Lorentz group $\rL={\rm SO}^0(1,3)$. As is well-known (\cite{green},
p.27), $L$ is homeomorphic with ${\bf RP}^3\times {\bf R}^3$. The spin group
$L_{\rs}$ acts on the spinor space $V$ by the generators
\beq
L_{ab}=\frac{1}{4}[\g_a,\g_b]. \label{b3213}
\eeq
By virtue of the relation
\be
L_{ab}^+\g^0=- \g^0L_{ab},
\ee
this action preserves the spinor metric (\ref{b3201}).
Accordingly, the Lorentz group L acts on the Minkowski space $M$ by the
generators
\beq
L_{ab}{}^c{}_d= \eta_{ad}\dl^c_b- \eta_{bd}\dl^c_a. \label{b3278}
\eeq

\begin{rem}
The generating elements $e\in M$, $\eta(e,e)=\pm 1$, of
the group Pin$(1,3)$ act on the Minkowski space by the adjoint
representation which is the composition 
\be
e: v\to eve^{-1} =-v+2\frac{\eta(e,v)}{\eta(e,e)}e, \qquad e,v\in{\bf
R}^4, 
\ee
of the total reflection of $M$ and the reflection across the hyperplane
\be
e^\perp=\{w\in M: \eta(e,w)=0\}
\ee
which is perpendicular to $e$ with respect to the metric $\eta$ in $M$. By 
the well-known Cartan--Dieudonn\'e theorem, every element of the
pseudo-orthogonal
group O$(p,q)$ can be written as a product of $r\leq p+q$ reflections
across hyperplanes in the vector space ${\bf R}^{p+q}$ (\cite{law}, p.17).
In particular, the group Spin$(1,3)$ consists of the elements of Pin$(1,3)$
which comprise an even number of reflections. The epimorphism of
Spin$(1,3)$ onto the Lorentz group SO$(3,1)$, as like as the epimorphism
(\ref{b3204}), are determined by the fact that the elements
$e^\perp$ and $(-e)^\perp$ define the same reflection.
\end{rem}

Let us now consider a bundle of Clifford algebras 
${\bf C}_{1,3}$ over a world
manifold $X$ whose 
structure group is the spin group $L_\rs$ of 
automorphisms of ${\bf C}_{1,3}$.
It possesses as subbundles the spinor bundle 
$S_M\to X$, associated with a $L_\rs$-principal bundle $P_\rs$, 
and a bundle $E_M\to X$ of Minkowski spaces of generating elements of 
${\bf C}_{3,1}$, which is associated with the L-principal bundle $P_\rs/{\bf
Z}_2$. 

The bundle
$E_M\to X$ of Minkowski spaces must be isomorphic to the cotangent bundle
$T^*X$ in order that sections of the spinor bundle $S_M$ describe Dirac's
fermion fields on a world manifold
$X$. In other words, we must consider a spin structure on the cotangent
bundle $T^*X$ of $X$ \cite{law}.

There are several almost equivalent definitions of a spin structure
\cite{avis,fat95,fat95a,heu,law}. One can say that a 
pseudo-Riemannian spin structure 
 on a world manifold $X$ is a pair
$(P_\rs, z_\rs)$ of an $L_\rs$-principal bundle $P_\rs\to X$ and a principal
bundle morphism
\beq
z_\rs: P_\rs \to LX \label{b3246}
\eeq
over $X$ of $P_s$
to the $GL_4$-principal frame
bundle
$LX\to X$. More generally, one can define a spin structure on any
vector bundle $E\to X$  (see \cite{law}, p.80). Then the precedent definition
applies to the particular case in which $E$ is the cotangent bundle
$T^*X$ and the fibre metric in $T^*X$ is a pseudo-Riemannian metric.

Since the homomorphism $L_\rs \to GL_4$
factorizes trough the epimorphism (\ref{b3204}), every bundle morphism
(\ref{b3246}) factorizes through a morphism of $P_\rs$ onto some
principal subbundle of the frame bundle $LX$ with the proper Lorentz 
group L as a structure group. It follows that the necessary condition for the
existence of a Lorentz spin structure on $X$ is that 
the structure group $GL_4$ of 
$LX$ is reducible to the proper Lorentz group L. 

From the physical viewpoint, it means that the
existence of Dirac's fermion matter implies the
existence of a gravitational field.

\section{Reduced structure}

First, we recall some general notions. Let 
\be
\pi_{PX}:P\to X
\ee
be a principal bundle with a structure group $G$, which acts freely and
transitively on the fibres of $P$ on the right:
\beq
R_g : p\mapsto pg, \quad  p\in P,\quad g\in G. \label{b1}
\eeq

Let $Y$ be a $P$-associated fibre bundle with 
a standard fibre $V$ on which the
structure group $G$ of $P$ acts on the left. Recall that
all associated fibre bundles with the same typical fibre 
are isomorphic to each
other, but the isomorphisms are not canonical in general. Unless
otherwise stated, we restrict our considerations to the 
canonically associated
fibre bundles, which are the quotient
\beq
Y=(P\times V)/G \label{b3103}
\eeq
with respect to the identification of elements $(p,v)$ and $(pg,g^{-1}v)$,
$g\in G$.  Let this fibre bundle be  coordinatized by $(x^\la, y^i)$.
Automorphisms of the principal bundle $P$, bundle atlases and principal
connections on $P$ define similar objects on the associated fibre bundles
(\ref{b3103}) in a canonical way, in contrast with other
associated fibre bundles.

\begin{rem}
Let $[p]$ denote
the restriction of the canonical morphism
\be
P\times V\to (P\times V)/G
\ee
to $p\times V$. For the sake of simplicity, we  shall write 
\be
[p](v)= [p,v]_G, \qquad p\in P,\ v\in V.
\ee
\end{rem}

By a principal automorphism of a principal bundle $P$ is
meant an automorphism $\Phi_P$ which is
equivariant under the canonical action (\ref{b1}), that is, the diagram
\be
\begin{array}{rcccl}
 & {P} &  \op\longrightarrow^{R_g} & {P} &  \\
{{}_{\Phi_P}} &\put(0,10){\vector(0,-1){20}} & 
& \put(0,10){\vector(0,-1){20}} &
{{}_{\Phi_P}} \\
 & {P} & \op\longrightarrow_{R_g} & {P} & 
\end{array}
\ee 
is commutative. A principal automorphism yields the corresponding
automorphism 
\beq
\Phi_Y: (P\times V)/G\to  (\Phi_P(P)\times V)/G \label{024}
\eeq
of every fibre bundle $Y$ associated with $P$.

Every principal automorphism of a principal bundle $P$ is represented as
\beq
\Phi_P(p)=pf(p), \qquad p\in P, \label{b3111}
\eeq
where $f$ is a $G$-valued equivariant function on $P$, i.e.,
\beq
f(pg)=g^{-1}f(p)g, \qquad \forall g\in G. \label{b3115}
\eeq
There is a one-to-one correspondence
between the
functions $f$ (\ref{b3115}) and the global sections
$s$ of the $P$-associated group bundle
\beq
P^G =(P\times G)/G, \label{b3130}
\eeq
whose typical fibre is the group $G$ itself and the action is given by the
adjoint representation (\cite{guil}, p.277). There is defined the canonical
fibre-to-fibre action of the group bundle $P^G$ on any
$P$-associated bundle $Y$: 
\ben
&& P^G\op\times_X Y\to Y, \nonumber \\
&& ([p, g]_G, [p, v]_G) \to [p, gv]_G, \qquad g\in G,\ v\in V.
\label{b3131'}
\een
Then, the above mentioned correspondence is given by the relation
\be
pf(p)=s(\pi_{PX}(p))p.
\ee

Let $H$ be a  
closed subgroup of $G$. We have the composite fibre bundle
\beq
P\to P/H\to X, \label{b3223a}
\eeq
where
\beq
(i)J\quad \Si:=P/H\ar^{\pi_{\Si X}} X \label{b3193}
\eeq
is a $P$-associated fibre bundle with the typical fibre $G/H$ on which the
structure group $G$ acts naturally on the left, and 
\beq
(ii) \quad P_\Si:=P\ar^{\pi_{P\Si}} P/H \label{b3194}
\eeq
is a principal bundle with the structure group $H$ (\cite{kob}, p.57).

One says that the structure group $G$ of a principal bundle $P$ is reducible
to the subgroup
$H$ if there exists a principal
subbundle
$P^h$ of $P$ with the structure group $H$. This subbundle is called the 
reduced $G^\downarrow H$-structure 
 \cite{gor,kob72,zul}.

\begin{rem}
Note that in \cite{gor} and \cite{kob72} the authors 
are concerned with reduced
structures on the principal frame bundle
$LX$. This notion is generalized to arbitrary
principal bundle in \cite{zul}. In
\cite{gor} a reduced structure  is regarded as a monomorphism of a given
principal bundle $P\to X$ with a structure group $H$ into the principal frame
bundle $LX$. Thereby, GL$(4,{\bf R})^\downarrow H$-structures are defined up
to isomorphisms.
\end{rem}

Let us recall the following theorems. 
 
\begin{theo}
A structure group
$G$ of a principal bundle $P$ is reducible to a closed subgroup $H$ iff $P$
has an atlas $\Psi$ with $H$-valued transition functions (\cite{kob},
p.53). 
\end{theo}

Given a reduced
subbundle $P^h$ of $P$, such an atlas $\Psi$ is defined by a family of local
sections $\{z_\al\}$ which take their values into $P^h$.

\begin{theo}\label{redsub}
There is a one-to-one correspondence 
\be
P^h=\pi_{P\Si}^{-1}(h(X))
\ee
between the
 reduced $H$-principal subbundles 
$P^h$ of $P$ and the global sections $h$ of the quotient fibre bundle
$P/H\to X$ (\cite{kob}, p.57).
\end{theo}

 Given such a section $h$, let us consider 
the restriction $h^*P_\Si$ of the
$H$-principal bundle
$P_\Si$ (\ref{b3194}) to $h(X)$. This is a $H$-principal bundle over $X$
(\cite{kob}, p.60), which is isomorphic to the reduced
subbundle
$P^h$ of $P$.   

In general, there are topological obstructions to the reduction of a
structure group of a principal bundle to a subgroup. In accordance with
a well-known theorem (see \cite{ste}, p.53), a structure group $G$ of a
principal bundle $P$ is always reducible to a closed subgroup $H$ if the
quotient
$G/H$ is homeomorphic to an Euclidean space ${\bf R}^k$ (\cite{ste}, p.53). 
In this case, all
$H$-principal subbundles of $P$ are isomorphic to each
other as
$H$-principal bundles (\cite{ste}, p.56). 

In particular, a structure group $G$ of a principal 
bundle is always reducible
to its maximal compact subgroup $H$ since the quotient space $G/H$ is
homeomorphic to an Euclidean space (\cite{ste}, p.59). It follows that there
is a one-to-one correspondence between equivalence classes of $G$-principal
bundles and those of $H$-principal bundles if $H$ is a maximal compact
subgroup of $G$ \cite{hir,ste}. In particular, this consideration applies
to GL$(n, {\bf R})$- and O$(n)$-principal bundles as like as GL$^+(n, {\bf
R})$- and SO$(n)$-principal bundles.

\begin{prop}\label{isomorp1}
Every vertical principal automorphism $\Phi$ of the
principal bundle
$P\to X$ sends a reduced subbundle $P^h$ onto an isomorphic $H$-principal
subbundle
$P^{h'}$. 
\end{prop}

\begin{proof}
Let 
\be
\Psi^h=\{(U_\al,z^h_\al), \rho^h_{\al\bt}\}, \qquad z^h_\al(x) =
z^h_\bt(x)\rho^h_{\al\bt}(x), \qquad x\in U_\al\cap U_\bt,
\ee
be an atlas of the reduced
subbundle $P^h$, where $z^h_\al$ are local sections of $P^h\to X$  and
$\rho^h_{\al\bt}$ are the transition functions. Given a vertical automorphism
$\Phi$ of $P$, let us provide the reduced 
subbundle $P^{h'}=\Phi(P^h)$ with the
atlas 
\be
\Psi^{h'}=\{(U_\al,z^{h'}_\al), \rho^{h'}_{\al\bt}\}
\ee
determined by the local
sections
\be
z^{h'}_\al =\Phi\circ z^h_\al
\ee
of $P^{h'}\to X$. Then it is readily observed that 
\be
\rho^{h'}_{\al\bt}(x) =\rho^h_{\al\bt}(x), \qquad x\in U_\al\cap U_\bt.
\ee
\end{proof}

\begin{prop}\label{isomorp2}
Conversely, let two reduced subbundles $P^h$ and $P^{h'}$ of a principal
bundle $P$ be isomorphic to each other as $H$-principal bundles and
let $\Phi:P^h\to P^{h'}$ be an isomorphism. Then $\Phi$ can be extended to 
a vertical automorphism of $P$. 
\end{prop}

\begin{proof}
The isomorphism $\Phi$ determines a
$G$-valued  function $f$ on $P^h$ given by the relation 
\be
pf(p)=\Phi(p), \qquad p\in P^h.
\ee
Obviously, this function is $H$-equivariant. Its prolongation to a
$G$-equivariant function on $P$ is defined to be
\be
f(pg)=g^{-1}f(p)g, \qquad p\in P^h, \qquad g\in G.
\ee
In accordance with the
relation (\ref{b3111}), this function defines a principal automorphism of $P$
whose restrction to $P^h$ coincides with $\F$.
\end{proof}

Given a reduced subbundle $P^h$ of a principal bundle $P$, let
\be
Y^h=(P^h\times V)/H
\ee
be the canonically associated fibre bundle with a typical fibre $V$.
If $P^{h'}$ is another reduced subbundle of $P$ which is isomorphic to $P^h$,
the fibre bundle $Y^h$ is associated with $P^{h'}$, but not canonically
associated  in general.
One can write
\be
&& Y^h\ni [p, v]_H= [pf(p), f(p)^{-1}v]_H\in Y^{h'},\\
&& p\in P^h,
\qquad pf(p)\in P^{h'}, \qquad v\in V,
\ee
if a typical fibre $V$ of $Y^h$ admits
representation of the whole group
$G$ only.

\section{Tetrad fields}

Since a world manifold is assumed to be parallelizable, the structure group
$GL_4$ of the frame bundle $LX$ is obviously reducible to the Lorentz group L.
The subbundle $L^hX$ is said to be a Lorentz structure.

In accordance with Theorem \ref{redsub}, there is
a one-to-one correspondence between the reduced L-principal subbundles
$L^hX$ of
$LX$ and the global
sections $h$ of the quotient fibre bundle 
\beq
\Si_{\rm T}=LX/\rL, \label{5.15}
\eeq
called the  tetrad bundle. 
This  is a $LX$-associated fibre bundle with the typical 
fibre
$GL_4/L$. Since the group $GL_4$ is homotopic to its maximal compact subgroup
SO$(4)$ and the proper Lorentz group is homotopic to its maximal compact
subgroup SO$(3)$, $GL_4/L$ is homotopic to the Stiefel manifold
\be
{\bf V}(4,1;{\bf R})=SO(4)/SO(3)=S^3
\ee
(\cite{ste}, p.33), and this is homeomorphic to the topological space
$S^3\times{\bf R}^7$. The fibre bundle (\ref{5.15}) is the 2-fold covering of
the metric bundle
$\Si_{\rm PR}$ (\ref{b3203}). Its global sections are called tetrad
fields.

Since $X$ is parallelizable, any two Lorentz subbundles $L^hX$ and $L^{h'}X$
are isomorphic to each other. It follows that, in virtue of Proposition
\ref{isomorp2}, there exists a vertical bundle automorphism
$\Phi$ of $LX$ which sends $L^hX$ onto $L^{h'}X$.
The associated  vertical automorphism $\Phi_\Si$ of the fibre bundle
$\Si_{\rm T}\to X$ transforms the tetrad field $h$ to the tetrad field $h'$.

Every tetrad field $h$ defines an associated  Lorentz atlas
$\Psi^h=\{(U_\zeta,z_\zeta^h)\}$ of 
$LX$ such that the corresponding local sections $z_\zeta^h$ of the frame
bundle $LX$ take their values into its Lorentz subbundle $L^hX$. 

Given a Lorentz atlas $\Psi^h$, the
pull-back 
\beq
h^a\ot t_a=z_\zeta^{h*}\th_{LX}=h_\la^a dx^\la\ot t_a \label{b3211}
\eeq
of the canonical form $\th_{LX}$ (\ref{b3133'}) by a local
section
$z_\zeta^h$ is said to be a (local)  tetrad form. 
The tetrad form (\ref{b3211}) determines the  tetrad
coframes
\be
t^a(x) = h^a_\m(x)dx^\m, \qquad x\in U_\zeta,
\ee
in the cotangent bundle $T^*X$, which are denoted by the same symbol
$h^a$ for the sake of simplicity. These coframes are associated with the
Lorentz atlas $\Psi^h$.

The coefficients $h^a_\m$ of the tetrad forms and the inverse matrix elements
\beq
h^\m_a =H^\m_a\circ z^h_\zeta. \label{L6}
\eeq
are called the  tetrad functions.  
Given a Lorentz atlas $\Psi^h$, the tetrad field $h$ can be represented by the
family of tetrad functions $\{h^\m_a\}$.
In particular, we have the well-known
relation 
\be
&& g=h^a\ot h^b\eta_{ab}, \\
&& g_{\m\nu}=h_\m^ah_\nu^b\eta^{ab},
\ee 
between tetrad functions and metric functions of the corresponding
pseudo-Riemannian metric $g:X\to\Si_{\rm PR}$. 

\begin{rem}\label{spacetime}
Since the world manifold $X$ is assumed to be parallelizable, 
it admits global tetrad
forms (\ref{b3211}). They, however, are not canonical. 
 
Even if $X$ were not parellelizable, the existence of a Lorentz
structure guarantees that there is a Lorentz atlas such that the temporal
tetrad form
$h^0$ is globally defined. This is a consequence of the fact that $L$ is
reducible to its maximal
compact subgroup SO$(3)$ and, therefore, there exists a
SO$(3)$-principal subbundle $L^h_0X\subset L^hX$, called a  space-time
structure associated with $h$. The corresponding global section of the
quotient fibre bundle 
$L^hX/{\rm SO}(3)\to X$ with the typical fibre ${\bf R}^3$ is a 
3-dimensional  spatial distribution
$FX\subset TX$ on $X$. Its generating 1-form
written relative to a Lorentz
atlas is exactly the global tetrad form $h^0$ \cite{sardz92}. There is the
corresponding decomposition
\be
TX=FX\oplus NF,
\ee
where $NF$ is the 1-dimensional fibre bundle defined by the  tetrad frame
 $h_0=h^\m_0\dr_\m$. This decomposition is called a
 space-time decomposition.  In
particular, if the generating form $h^0$ is exact, the space-time
decomposition obeys the condition of stable causality by Hawking \cite{haw}. 
\end{rem}

\section{Pseudo-Riemannian spin structure}

Given a tetrad field $h$, let $L^hX$ be the corresponding reduced Lorentz
subbundle. Since $X$ is non-compact and parallelizable, the principal bundle
$L^hX$ is extended uniquely (with accuracy of an automorphism) to a
$L_\rs$-principal bundle $P^h\to X$ \cite{ger} and to a fibre bundle of
Clifford algebras ${\bf C}_{1,3}$ (\cite{law}, p.95). 
There is the principal bundle morphism 
\beq
z_h: P^h \to L^hX \subset LX \label{b3195}
\eeq
over $X$ such that
\be
z_h\circ R_g =R_{z_L(g)}, \qquad \forall g\in L_s.
\ee
This is a $h$-associated pseudo-Riemannian spin structure 
on a world manifold.
We call $P^h$ the  $h$-associated principal spinor bundle. 
Every Lorentz
atlas
$\Psi^h=\{z^h_\zeta\}$ of $L^hX$ can be lifted to an atlas of the principal
spinor bundle $P^h$.

Let us consider the corresponding $L^hX$-associated fibre bundle $E_M$ of
Minkowski spaces
\beq
E_M=(L^hX\times M)/\rL=(P^h\times M)/L_s \label{b3192}
\eeq
and the $P^h$-associated spinor bundle
\beq
\bx{S^h=(P^h\times V)/L_s,}\label{510}
\eeq
called in sequel the  $h$-associated
spinor bundle. 
The fibre bundle 
$E_M$ (\ref{b3192}) is isomorphic to the cotangent
bundle 
\be
\bx{T^*X=(L^hX\times M)/\rL}
\ee
as a fibre bundle with the structure Lorentz group L.
Then there exists the representation
\beq
\bx{\g_h: T^*X\ot S^h=(P^h\times (M\ot V))/L_s\to (P^h\times
\g(M\ot V))/L_s=S^h} \label{L4}
\eeq
of covectors to $X$ by the Dirac $\g$-matrices
on elements of the spinor bundle $S^h$. 

Relative to an atlas $\{z_\zeta\}$ of $P^h$ and to 
the associated Lorentz atlas
$\{z^h_\zeta=z_h\circ z_\zeta\}$ of $LX$, the
representation (\ref{L4}) reads
\be
y^A(\g_h(h^a(x) \ot v))=\g^{aA}{}_By^B(v), \qquad v\in S^h_x,
\ee
where $y^A$ are the
corresponding bundle coordinates of $S^h$ and $h^a$ are the tetrad forms
(\ref{b3211}). 
As a shorthand, we can write
\be
&& \wh h^a=\g_h(h^a)=\g^a,\\
&&\wh dx^\la=\g_h(dx^\la)=h^\la_a(x)\g^a.
\ee

Sections $s_h$ of the $h$-associated spinor 
bundle $S^h$ (\ref{510}) describe
Dirac's fermion fields in the presence of the tetrad field $h$. Indeed,
let $A_h$ be a principal connection on $S^h$ and
\be
&&D: J^1S^h\to T^*X\op\ot_{S^h} S^h,\\
&&D=(y^A_\la-A^{ab}{}_\la L_{ab}{}^A{}_By^B)dx^\la\ot\dr_A,
\ee
the corresponding covariant differential. Here, we have used the isomorphism
\be
VS^h\cong S^h\op\times_XS^h.
\ee
 The first order
differential  Dirac operator is defined on $S^h$
as the composition 
\ben
&&\bx{\cD_h=\g_h\circ D: J^1S^h\to T^*X\ot S^h\to S^h,} \label{l13}\\
&& y^A\circ\cD_h=h^\la_a(\g^a)^A{}_B(y^B_\la- \frac12 A^{ab}{}_\la
L_{ab}{}^A{}_By^B). \nonumber
\een

\begin{rem}
The spinor bundle $S^h$ is a complex fibre bundle with a real structure group
over a real manifold. One can regard such a fibre bundle  as
a real one whose real dimension of the fibres is equal to the doubled complex
dimension. In particular, the jet manifold $J^1S^h$
of $S^h$ is defined as usual. It is coordinatized by  $(x^\la,y^A,y^A_\la)$.
\end{rem}

The $h$-associated spinor bundle $S^h$ is equipped with the  fibre spinor
metric 
\ben
&& a_h: S^h\op\times_X S^h \to {\bf R}, \label{b3214} \\
&& a_h(v,v')=\frac12(v^+\g^0v' +{v'}^+\g^0v), \qquad v,v'\in S^h.\nonumber
\een
Using this metric and the Dirac operator (\ref{l13}), one can define the
 Dirac Lagrangian density on $J^1S^h$ in the
presence of a background tetrad field $h$ and a background connection $A_h$
on $S^h$ as
\be
&& L_h:J^1S^h\to \op\w^4T^*X, \\
&& L_h=[a_h(i\cD_h(w),w) -ma_h(w,w)]h^0\w \cdots\w h^3, \qquad w\in J^1S^h.
\ee
Its coordinate expression is
\ben
&& \cL_h=\{\frac{i}{2}h^\la_q[y^+_A(\g^0\g^q)^A{}_B(y^B_\la
-\frac12A_\la{}^{ab} L_{ab}{}^B{}_Cy^C) - \label{b3215}\\
&& \qquad (y^+_{\la A} -\frac12A_\la{}^{ab}y^+_CL_{ab}^+)
(\g^0\g^q)^A{}_By^B] - my^+_A(\g^0)^A{}_By^B\}\det(h^a_\m). \nonumber
\een

\section{Spin connections}

Note that there is a one-to-one correspondence between the principal
(spin) connections on the
$h$-associated principal spinor bundle
$P^h$ and the principal (Lorentz) connections
 on the L-principal bundle
$L^hX$ as follows.

First, let us recall the following theorem (\cite{kob},
p.79). 

\begin{theo}\label{connmorp}
Let $P'\to X$ and $P\to X$ be principle bundles with the structure groups
$G'$ and $G$, respectively. If $\Phi: P'\to P$ is a principal bundle morphism
over
$X$ with the corresponding homomorphism $G'\to G$, there exists a unique
principal connection
$A$ on $P$ for a given principal connection $A'$ on $P'$ such that $T\Phi$
sends the horizontal subspaces of $A'$ onto horizontal subspaces of $A$ 
\end{theo}

It follows that every principal connection 
\beq
A_h=dx^\la\ot(\dr_\la + \frac12A_\la{}^{ab} e_{ab}) 
\label{b3205}
\eeq
on $P^h$ defines a principal connection on
$L^hX$ which is given by the same expression (\ref{b3205}).
Conversely, the pull-back $z_h^*\om_A$ on
$P^h$ of a connection form $\om_A$ of a Lorentz connection $A_h$ on $L^hX$ is
equivariant under the the action of group
$L_\rs$ on
$P^h$, and this is a connection form of a spin connection on $P^h$. 

In particular, as is well-known, the Levi--Civita connection of a
pseudo-Riemannian metric
$g$ gives rise to a spin connection with the components
\beq
A_\la{}^{ab}=\eta^{kb}h^a_\m(\dr_\la h^\m_k - h^\nu_k\{_\la{}^\m{}_\nu\})
\label{b3217}
\eeq
on the $g$-associated spinor
bundle $S^g$.

In gauge gravitation theory, Lorentz connections are treated as gauge
potentials associated with the Lorentz group.  
At the same time, every world connection $K$ on a world manifold $X$ also
defines a spin connection on a $h$-associated principal spinor bundle
$P^h$. It follows that the gauge gravitation theory  is reduced to
metric-affine gravitation theory in the presence of Dirac's fermion fields
\cite{giach97,sard97b}.

Note that, in accordance with Theorem \ref{connmorp}, every Lorentz
connection $A_h$ (\ref{b3205})
on a reduced Lorentz subbundle $L^hX$ of $LX$ gives rise to a world
connection $K$ (\ref{08}) on $LX$ where
\beq
K_\la{}^\m{}_\nu = h^k_\nu\dr_\la h^\m_k + \eta_{ka}h^\m_b h^k_\nu
A_\la{}^{ab}. \label{b3207}
\eeq
At the same time, every principal connection $K$ on 
the $GL_4$-principal bundle
$LX$ defines a Lorentz principal connection $K_h$ on 
a $L$-principal subbundle
$L^hX$ as follows.

It is readily observed that the Lie algebra of the general linear group
$GL_4$ is the direct sum
\be
{\bf g}(GL_4) = {\bf g}(\rL) \oplus {\bf m}
\ee
of the Lie algebra ${\bf g}(\rL)$ of the Lorentz group and a subspace ${\bf
m}\subset {\bf g}(GL_4)$ such that 
\be
ad(l)({\bf m})\subset {\bf m}, \qquad l\in L
\ee
where $ad$ is the adjoint representation. Let $\om_K$ 
be a connection form of a
world connection $K$ on $LX$. Then, by the well-known theorem (\cite{kob},
p.83), the pull-back onto $L^hX$ of the ${\bf g}(\rL)$-valued 
component $\om_L$
of
$\om_K$ is a connection form of a principal connection $K_h$ on the reduced
Lorentz subbundle $L^hX$. To obtain the
connection parameters of $K_h$, 
let us consider the local connection 1-form of the connection $K$ with
respect to a Lorentz atlas $\Psi^h$ of $LX$ given by the tetrad forms
$h^a$. This reads
\be
&& {z^h}^*\om_K= K_\la{}^b{}_k dx^\la\ot e_b{}^k,\\
&& K_\la{}^b{}_k = -h^b_\m \dr_\la h^\m_k  + K_\la{}^\m{}_\nu h^b_\m
h^\nu_k,
\ee
where $\{e_b{}^k\}$ is the basis of the right Lie algebra of the group $GL_4$.
Then, the Lorentz part of this form is accurately the
local connection 1-form of the connection $K_h$ on $L^hX$. We have
\ben
&&z^{h*}\om_L= \frac12 A_\la{}^{ab}dx^\la\ot e_{ab}, \label{K102} \\
&& A_\la{}^{ab} =\frac12 (\eta^{kb}h^a_\m-\eta^{ka}h^b_\m)(\dr_\la h^\m_k -
 h^\nu_k K_\la{}^\m{}_\nu). \nonumber 
\een
If $K$ is a Lorentz connection $A_h$, then obviously $K_h=A_h$.

Accordingly, the connection $K_h$ on $L^hX$ given by the local connection
1-form (\ref{K102}) defines the corresponding spin connection on $S^h$
\beq
\bx{K_h=dx^\la\ot[\dr_\la +\frac14 (\eta^{kb}h^a_\m-\eta^{ka}h^b_\m)(\dr_\la
h^\m_k - h^\nu_k K_\la{}^\m{}_\nu)L_{ab}{}^A{}_B y^B\dr_A],} \label{b3212}
\eeq
where $L_{ab}$ are the generators (\ref{b3213})  
\cite{giach97,sard97b}. Such a connection has been considered in
\cite{ar,pon,tuc}

A substitution of the spin connection (\ref{b3212}) into the Dirac
operator (\ref{l13}) and into the Dirac Lagrangian density (\ref{b3215})
provides a description of Dirac's fermion fields in the presence of arbitrary
linear connections on a world manifold, not only 
in the presenze of the Lorentz
ones.

One can utilize the connection (\ref{b3212}) for constructing a horizontal
lift of a vector field $\tau$ on $X$ onto $S^h$. This lift reads
\beq
\tau_{K_h} = \tau^\la\dr_\la +\frac14
\tau^\la(\eta^{kb}h^a_\m-\eta^{ka}h^b_\m) (\dr_\la
h^\m_k - h^\nu_k K_\la{}^\m{}_\nu) L_{ab}{}^A{}_B y^B\dr_A. \label{b3218}
\eeq
For every vector field $\tau$ on $X$, let us 
choose a symmetric connection $K$
which has $\tau$ as a geodesic vector field. Then we get the canonical
horizontal lift 
\beq
\bx{\wt\tau = \tau^\la\dr_\la +\frac14
(\eta^{kb}h^a_\m-\eta^{ka}h^b_\m) (\tau^\la\dr_\la
h^\m_k - h^\nu_k\dr_\nu\tau^\m) L_{ab}{}^A{}_B y^B\dr_A} \label{b3216}
\eeq
of vector fields $\tau$ on $X$ onto the $h$-associated spinor bundle $S^h$
\cite{sard97b}.

\begin{rem}
The canonical lift (\ref{b3216}) is brought into the form
\be
\wt\tau = \tau_{\{\}} - \frac14
(\eta^{kb}h^a_\m-\eta^{ka}h^b_\m)h^\nu_k\nabla_\nu\tau^\m L_{ab}{}^A{}_B
y^B\dr_A,
\ee
where $\tau_{\{\}}$ is the horizontal lift (\ref{b3218}) of $\tau$ by means
of the spin Levi--Civita connection (\ref{b3217}) 
of the tetrad field $h$, and
$\nabla_\nu \tau^\m$ are the covariant derivatives of $\tau$ relative to the
same Levi--Civita connection. This is exactly the Lie derivative of
spinor fields described in \cite{fat95,koss}.
\end{rem}

\section{Universal spin structure}

The canonical lift  (\ref{b3216}) fails to be a
generator of general covariant transformations because it does not involve
transformations of tetrad fields. To define general covariant transformations
of spinor bundles, we should consider spinor structures associated with
different tetrad fields. The
difficulty arises because, though the principal spinor bundles $P^h$ and
$P^{h'}$ are isomorphic, the
$h$-associated spinor bundle
$S^h$ fails to be 
$h'$-associated since it is associated, but not canonically associated
with $P^{h'}$. As a consequence, the
representations
$\g_h$ and $\g_{h'}$ (\ref{L4}) for different tetrad fields $h$ and $h'$ 
are not equivalent \cite{sardz92,sard95}. 

Indeed, let $L^hX$ and $L^{h'}X$ be the Lorentz subbundles of the frame
bundle $LX$ and $\Phi$ an automorphism of $LX$, characterized by an
equivariant function $f$ on $LX$, which sends $L^hX$ onto $L^{h'}X$. Let
\be
&&t^*=[p, e]_{\rm L}=[pf(p), (f^{-1}(p)e]_{\rm L}, \qquad p\in L^hX,
\qquad e\in {\bf R}^4, \\
&& t^*=t_\m dx^\m=t_ah^a=t'_a{h'}^a,
\ee
be an element of $T^*X$. Its representations $\g_h$ and $\g_{h'}$ (\ref{L4})
read
\be
&& \g_h(t^*)=\g(e)=t_\m h^\m_a\g^a, \\
&& \g_{h'}(t^*)=\g(f^{-1}(p)e)=t_\m {h'}^\m_a{\g'}^a.
\ee
These representations are not equivalent since 
no isomorphism $\Phi_\rs$ of $S^h$ onto $S^{h'}$ can obey the
condition
\be
\g_{h'}(t^*)=\Phi_\rs \g_h(t^*)\Phi_\rs^{-1}, \qquad \forall t^*\in T^*X.
\ee

It follows that every Dirac's fermion field must be described 
by a pair with a
certain tetrad (gravitational) field.  Thus we observe the phenomenon of
symmetry breaking in gauge gravitation theory which exhibits the physical
nature of gravity as a Higgs field \cite{sardz92}. 
The goal is to describe the totality of fermion-gravitation pairs.

\begin{rem}
All spin structures on a manifold $X$ which
are related to the two-fold universal covering groups possess the following
two properties \cite{greub}.

Let $P\to X$ be a principal bundle with a structure group $G$ with the
fundamental group $\pi_1(G)={\bf Z}_2$. Let $\wt G$ be the universal covering
group of $G$.

1. The topological obstruction for the existence of a $\wt G$-principal bundle
$\wt P\to X$ covering the bundle $P$ is a  non-zero element of the \v Cech
cohomology group
$H^2(X;{\bf Z}_2)$ of $X$ with coefficients in ${\bf Z}_2$.

2. Inequivalent lifts of $G$-principal bundle $P$ 
to a $\wt G$-principal bundle
are classified by elements of the \v Cech cohomology group $H^1(X;{\bf Z}_2)$.

In particular, the well-known topological obstruction in order that a
Riemannian spin structure and a pseudo-Riemannian spin structure
can exist on $X$ is the non-zero  second Stiefel--Whitney  class 
$w_2(X)\in H^2(X;{\bf Z}_2)$ of $X$ (\cite{law}, p.82).
The set of these inequivalent spin structures is 
in bijective correspondence with the cohomology group
$H^1(X;{\bf Z}_2)$
(\cite{greub,swit}; \cite{law}, p.82). 
In the case of 4-dimensional noncompact manifolds that we consider,
all Riemannian and pseudo-Riemannian spin structures are equivalent
\cite{avis,ger}.
\end{rem}

\begin{ex}\label{riem} {\bf Riemannian spin structure.}
Let us consider spin structures on Riemannian manifolds.
 Let $X$ be an
arbitrary 4-dimensional oriented manifold. The structure group $GL_4$ of the
principal frame bundle $LX$ is reducible to its maximal compact
subgroup SO$(4)$ since the quotient $GL_4/{\rm SO}(4)$ is homeomorphic to the
Euclidean space ${\bf R}^{10}$. It follows that
a  Riemannian metric $g_R$,
 represented by a section of the quotient fibre bundle 
\beq
\Si_{\rm R}:= LX/{\rm SO}(4)\to X, \label{b3220}
\eeq 
always  exists on a manifold $X$. The corresponding
SO$(4)$-principal subbundle $L^gX$ is called a
Riemannian structure on a world manifold $X$.

Given two different Riemannian metrics $g_R$ and $g'_R$ on $X$, the
corresponding SO$(4)$-principal subbundles $L^gX$ and $L^{g'}X$ of $LX$ are
isomorphic as 
SO$(4)$-principal bundles since the group space $GL_4$ is homotopic to
SO$(4)$. 

To introduce a Riemannian spin structure, one can consider the
complex Clifford algebra ${\bf C}_4$ which is generated by
elements of the vector space ${\bf R}^4$ equipped with the Euclidean metric
\cite{budi,law}. The corresponding spinor space $V_E$ is a minimal
left ideal of
${\bf C}_4$ provided with a Hermitian bilinear form. The spin group 
Spin$(4)$ is the two-fold universal covering group of
the group SO$(4)$. This is isomorphic to SU$(2)\ot$SU$(2)$ (\cite{corn},
p.430).

Let us assume
that the second Stiefel--Whitney class $w_2(X)$ of $X$ 
vanishes. A  Riemannian spin structure
 on a manifold $X$ is defined to be a pair
 of a Spin$(4)$-principal bundle
$P_\rs\to X$ and a principal bundle morphism $z$ of $P_\rs$ to
$LX$. Since such a morphism factorizes through a bundle morphism 
\be
z_g: P_\rs\to L^gX
\ee 
for some Riemannian metric $g_R$, this spin structure is a
$g_R$-associated spin structure. 
We denote the corresponding $g_R$-associated principal 
spinor bundle by $P^g$.
All these bundles on a 4-dimensional manifold $X$ are isomorphic \cite{avis}.
Note that, although spin principal bundles $P^g$ and $P^{g'}$ for different
Riemannian metric
$g$ and $g'$ are isomorphic,
the $g$-associated spinor bundle
\be
S^g=(P^g\times V_E)/{\rm Spin}(4)
\ee
is not canonically associated
with $P^{g'}$.
\end{ex}

The group $GL_4$ is not simply-connected. Its first homotopy group is
\be
\pi_1(GL_4)= \pi_1({\rm SO}(4)) ={\bf Z}_2
\ee
(\cite{green}, p.27). Therefore, $GL_4$
admits the
universal two-fold covering group
$\wt{GL}_4$  such that the diagram
\beq
\begin{array}{ccc}
 \wt{GL}_4 & \longrightarrow &  GL_4 \\
 \put(0,-10){\vector(0,1){20}} & 
& \put(0,-10){\vector(0,1){20}}  \\
{\rm Spin}(4) & \longrightarrow & {\rm SO}(4) 
\end{array} \label{b3242}
\eeq
is commutative \cite{heh,law,perc,swit}. 

A  universal spin structure on $X$ is
defined to be a pair consisting of a
$\wt{GL}_4$-principal bundle
$\wt{LX}\to X$ and a principal bundle morphism over $X$
\beq
\wt z: \wt{LX} \to LX \label{b3247}
\eeq
\cite{dabr,perc,swit}. There is the commutative diagram 
\beq
\begin{array}{ccc}
 \wt{LX} & \ar^{\wt z} &  LX \\
 \put(0,-10){\vector(0,1){20}} & 
& \put(0,-10){\vector(0,1){20}}  \\
P^g & \ar^{z_g} & L^gX 
\end{array} \label{b3222}
\eeq 
for any Riemannian metric $g_R$ \cite{perc,swit}.

Since the group $\wt{GL}_4$ is homotopic to the
group Spin$(4)$,
there is a one-to-one
correspondence between inequivalent universal spin
structures and inequivalent Riemannian spin structures \cite{swit}. In our
case, all universal spin structures as like as the Riemannian ones are
equivalent.

Given a universal spin structure (\ref{b3247}), one can consider the lift of
 bundle automorphisms of the frame bundle $LX$ (e.g., general covariant
transformations) to automorphisms of the principal bundle $\wt{LX} \to X$
and the associated fibre bundles \cite{dabr}. 
Spinor representations of
the group $\wt{GL}_4$, however, are infinite-dimensional \cite{heh,nee}.
Elements of this representation are called  world spinors. 
The corresponding field theory has been already developed (see \cite{heh} and
references therein).

A different procedure is to consider the commutative 
diagram
\beq
\begin{array}{rcl}
 \wt{LX}  & \op\longrightarrow^{\wt z} &  LX \\
  & \searrow  \swarrow & \\ 
 & {\Si_{\rm R}} &  
\end{array} \label{b3249}
\eeq
and the composite fibre bundle
\be
\wt{LX}\to \Si_{\rm R} \to X. 
\ee
Then the restriction of the Spin$(4)$-principal bundle $\wt{LX}\to \Si_{\rm
R}$ to
$g_R(X)\subset \Si_{\rm R}$ is isomorphic to the $g_R$-associated principal
spinor bundle $P^g$. 
This is the reason why the spin structure
(\ref{b3247}) is called the universal spin structure. Accordingly, the
universal spin structure (\ref{b3247}) over the fibre bundle $\Si_{\rm R}$
(\ref{b3220}), which is given by the diagram (\ref{b3249}) is said to be the 
universal Riemannian spin structure. 

Let us consider the composite spinor bundle
\beq
S\ar^{\pi_{S\Si}} \Si_{\rm R}\to X, \label{b3223'}
\eeq
where $S\to \Si_{\rm R}$ is the spinor bundle associated with the
Spin$(4)$-principal bundle $\wt{LX}\to \Si_{\rm R}$. Then,
whenever $g_R$ is a Riemannian metric on $X$,  sections of the spinor bundle
$S^g$ associated with a principal spinor bundle $P^g$ as in the commutative
diagram (\ref{b3222}) are in bijective correspondence 
with the sections $s$ of
the composite spinor bundle (\ref{b3223'}) which are projected onto $g_R$,
that is, $\pi_{S\Si}\circ s=g_R$.  

In a similar way, the universal
pseudo-Riemannian spin structure can be introduced.

\begin{rem}
It should be emphasized that the total space $S$ of the spinor bundle
(\ref{b3223'}) has the structure of the fibre bundle which
is associated with the $\wt{GL}_4$-principal bundle $\wt{LX}\to
X$ and whose typical fibre is the quotient 
\beq
(\wt{GL}_4\times V_E)/{\rm Spin}(4) \label{b3236}
\eeq
by identification of the elements
\be
(\wt g,v) \simeq (a\wt g,a^{-1}v), 
\qquad \wt g\in \wt{GL}_4,\ v\in V_E,\ a\in
{\rm Spin}(4).
\ee
Then, every morphism of the quotient
(\ref{b3236}) into the spin representation space of the group $\wt{GL}_4$
yields the corresponding morphisms of the composite spinor bundle
(\ref{b3223'}) into the
$\wt{GL}_4$-associated bundle of world spinors.
\end{rem}

\section{Spontaneous symmetry breaking}

The construction above using composite fibre bundles illustrates the standard
description of spontaneous symmetry breaking in gauge theories where matter
fields admit only exact symmetry transformations \cite{sard92,sardhp}.

Spontaneous symmetry breaking is a quantum phenomenon. In classical field
theory, spontaneous symmetry breaking is modelled by classical Higgs fields.
In gauge theory on a principal bundle $P\to X$, the necessary condition for
 spontaneous symmetry breaking is the
reduction of the structure group
$G$ of this principal bundle to its closed subgroup $H$ of exact symmetries
\cite{iva,nik,tra}. Higgs
fields  are described by global sections
$h$ of the quotient fibre bundle $\Si$ (\ref{b3193}). 

In accordance with Theorem \ref{redsub}, the set of Higgs fields $h$ is in
bijective correspondence with the set of reduced $H$-principal subbundles
$P^h$ of $P$. Given such a subbundle $P^h$, let 
\beq
Y^h=(P^h\times V)/H \label{b3235}
\eeq
be the associated fibre bundle with a typical fibre $V$. Its sections
describe matter fields in the presence of the Higgs fields $h$.

If $V$ does not admit the action of the whole symmetry
group $G$, the fibre bundle $Y^h$ (\ref{b3235}) is not associated
canonically with other $H$-principal subbundles.
It follows that $V$-valued matter
fields can be represented by pairs with a 
certain Higgs field only. The goal is to describe the totality of these pars
$(s_h,h)$ for all Higgs fields.

Let us consider the composite fibre bundle (\ref{b3223a}) and the composite
fibre bundle 
\beq
Y\ar^{\pi_{Y\Si}} \Si\ar^{\pi_{\Si X}} X, \label{b3225}
\eeq
where $Y\to \Si$ is the fibre bundle 
\beq
Y=(P\times V)/H \label{b3224}
\eeq 
associated with the principal bundle $P_\Si$ (\ref{b3194}) with the structure
group $H$ which acts on the typical fibre $V$ of $Y$ on the left. Given a
global section $h$ of the fibre bundle $\Si\to X$ (\ref{b3194}), let
\beq
Y^h=(P^h\times V)/H \label{b3227}
\eeq 
be  a fibre bundle associated with the reduced $H$-principal
subbundle
$P^h$ of 
$P$. There is the canonical isomorphism
\be
i_h: Y^h=(P^h\times V)/H \to (h^*P\times V)/H
\ee
of $Y^h$ to the subbundle of $Y\to X$ which is the restriction
\be
h^*Y=(h^*P\times V)/H
\ee
of the fibre bundle $Y\to\Si$ to $h(X)\subset \Si$. We have 
\beq
i_h(Y^h)\cong \pi^{-1}_{Y\Si}(h(X)). \label{b3226}
\eeq

Then every global section $s_h$ of the fibre bundle $Y^h$ corresponds to the
global section $i_h\circ s_h$ of the composite fibre bundle (\ref{b3225}).
Conversely, every global section $s$ of the composite fibre bundle
(\ref{b3225}) which is projected onto a section $h=\pi_{Y\Si}\circ s$ of the
fibre bundle $\Si\to X$ takes its values into the subbundle $i_h(Y^h)\subset
Y$ in accordance with the relation (\ref{b3226}). Thus, there is a
ono-to-one correspondence between the sections of the fibre bundle $Y^h$
(\ref{b3227}) and the sections of the composite fibre bundle (\ref{b3225})
which cover the section $h$.

Thus, it is the composite fibre bundle (\ref{b3225}) whose sections describe
the above mentione totality of the pairs $(s_h, h)$ of matter fields and
Higgs fields in gauge theory with broken symmetries \cite{sard92,sard95}.

The feature of the dynamics of field systems on composite fibre bundles
consists in the following.

Let  $Y$ (\ref{b3225}) be a composite fibre bundle coordinatized by $(x^\la,
\si^m, y^i)$, where $(x^\la, \si^m)$ are bundle coordinates of the fibre
bundle $\Si\to X$.  Let 
\beq
A_\Si=dx^\la\ot(\dr_\la+ A^i_\la\dr_i)
+d\si^m\ot(\dr_m+A^i_m\dr_i) \label{b3228}
\eeq
be a principal connection on the fibre bundle $Y\to \Si$.
This connection defines the splitting
\be
&& VY=VY_\Si\op\oplus_Y (Y\op\times_\Si V\Si), \\
&& \dot y^i\dr_i + \dot\si^m\dr_m=
(\dot y^i -A^i_m\dot\si^m)\dr_i + \dot\si^m(\dr_m+A^i_m\dr_i).
\ee
Using this splitting, one can construct
the first order differential operator
\ben
&&\wt D:J^1Y\to T^*X\op\ot_Y VY_\Si,\nonumber\\
&&\wt D=dx^\la\ot(y^i_\la- A^i_\la -A^i_m\si^m_\la)\dr_i,\label{7.10}
\een
on the composite fibre
bundle $Y$. 

The operator (\ref{7.10}) posesses the following property. 
Given a global section $h$ of $\Si$, its restriction 
\ben
&&\wt D_h =\wt D\circ J^1i_h: J^1Y^h \to T^*X\ot VY^h, \label{b3260}\\
&& \wt D_h =dx^\la\ot(y^i_\la- A^i_\la -A^i_m\dr_\la h^m)\dr_i, \nonumber
\een
to $Y^h$ is exactly the familiar covariant differential relative to the 
principal connection
\be
A_h=dx^\la\ot[\dr_\la+(A^i_m\dr_\la h^m + A^i_\la)\dr_i]
\ee
on the fibre bundle $Y^h\to X$, which is induced by the principal connection
(\ref{b3228}) on the fibre bundle $Y\to \Si$ by the imbedding $i_h$
(\cite{kob}, p.81).

Thus,
we may utilize $\wt D$ in order to construct a Lagrangian density on the
jet manifold $J^1Y$ of a composite fibre bundle which factorizes through $\wt
D$, that is,
\be
L:J^1Y\op\to^{\wt D}T^*X\op\ot_YVY_\Si\to\op\w^nT^*X.
\ee

\begin{rem}\label{higgs1} 
The total space of the composite fibre bundle $Y\to X$ (\ref{b3225}) can be
represented as the quotient
of the product $P\times G\times V$
by identification of the elements
\be
(p,g,v)\simeq (pab,b^{-1}g, a^{-1}v), \qquad \forall a\in H, \qquad \forall
b\in G.
\ee
It follows that $Y$ has the structure of the $P$-associated bundle
\be
Y=(P\times (G\times V)/H)/G
\ee
with the structure group $G$ and the typical fibre $(G\times V)/H$ which is
the quotient
of the product $G\times V$ by identification of the elements
\be
(g,v)\simeq (ag,a^{-1}v), \qquad \forall  a\in H.
\ee
In particular, if the typical fibre $V$ of the composite fibre bundle $Y\to X$
admits the action of the group $G$, these two bundle structures on $Y$ are
equivalent. 
\end{rem}

\section{Universal pseudo-Riemannian spin structure}

Let us turn now to fermion fields in gauge gravitation theory. 
We are based on
the following two facts.

\begin{prop}\label{ferm1}
The L-principal bundle 
\beq
P_{\rm L}:=GL_4\to GL_4/{\rm L} \label{b3244}
\eeq
is trivial.
\end{prop}

\begin{proof}
In accordance with the classification theorem (\cite{ste}, p.99), 
a $G$-principal bundle over an $n$-dimensional sphere $S^n$ is trivial if the
homotopy group $\pi_{n-1}(G)$ is trivial. The base space $Z=GL_4/{\rm L}$ of
the principal bundle (\ref{b3244}) is homeomorphic to $S^3\times {\bf R}^7$.
Let us consider the inclusion $f_1$ of $S^3$ 
into $Z$, $f_1(p)=(p,0)$,  and the
pull-back  L-principal bundle $f_1^*P_{\rm L}\to S^3$. 
Since L is homeomorphic
to ${\bf RP}^3\times {\bf R}^3$ and
$\pi_2({\rm L})=0$, this bundle is trivial. Let $f_2$ be the projection of 
$Z$ onto $S^3$. Then the pull-back
$L$-principal bundle $f_2^*(f^*_1P_{\rm L})\to Z$ is also
trivial. Since the composition morphism $f_1\circ f_2$ of $Z$ into $Z$ is
homotopic to the identity morphism of $Z$, the bundle $f_2^*(f^*_1P_{\rm
L})\to Z$ is equivalent to the bundle $P_{\rm L}$ (\cite{ste}, p.53). It
follows that the bundle (\ref{b3244}) is trivial.
\end{proof}

\begin{prop}\label{ferm2}
As in (\ref{b3242}), we have the commutative diagram
\beq
\begin{array}{ccc}
 \wt{GL}_4 & \longrightarrow &  GL_4 \\
 \put(0,-10){\vector(0,1){20}} & 
& \put(0,-10){\vector(0,1){20}}  \\
L_\rs & \ar^{z_L} & {\rm L} 
\end{array} \label{b3243}
\eeq
\end{prop}

\begin{proof}
The restriction of the universal covering group $\wt{GL}_4\to GL_4$ to the
Lorentz group L$\subset GL_4$ is obviously a covering space of L.
Let us show that this is the universal covering space. Indeed, any
non-contractible cycle in
$GL_4$ belongs to some subgroup $SO(3)\subset GL_4$ and the restriction of 
the fibre bundle $\wt{GL}_4\to GL_4$ to $SO(3)\subset GL_4$ is the universal
covering of
$SO(3)$. Since the proper Lorentz group is homotopic to its maximal compact
subgroup $SO(3)$, its universal covering space belongs to $\wt{GL}_4$.
\end{proof}

Let us consider the universal spin structure
$\wt{LX}\to X$. This is unique since
$X$ is parallelizable. In virtue of Proposition
\ref{ferm2}, we have the commutative diagram
\beq
\begin{array}{ccc}
 \wt{LX} & \ar^{\wt z} &  LX \\
 \put(0,-10){\vector(0,1){20}} & 
& \put(0,-10){\vector(0,1){20}}  \\
P^h & \ar^{z_h} & L^hX 
\end{array} \label{b3245}
\eeq 
for any tetrad field $h$. It follows that the quotient $\wt{LX}/L_\rs$ is
exactly the quotient $\Si_{\rm T}$ (\ref{5.15}) so that there is the
commutative diagram
\beq
\begin{array}{rcl}
 \wt{LX}  & \op\longrightarrow^{\wt z} &  LX \\
  & \searrow  \swarrow & \\ 
 & {\Si_{\rm T}} &  
\end{array} \label{b3250}
\eeq

By analogy with the diagram (\ref{b3249}), the diagram (\ref{b3250}) is
said to be the  universal pseudo-Riemannian spin
structure.  We have the composite fibre bundle
\beq
\wt{LX}\to \Si_{\rm T} \to X, \label{b3248}
\eeq
where $\wt{LX}\to \Si_{\rm T}$ is the $L_\rs$-principal bundle.

The universal pseudo-Riemannian spin structure \ref{b3250} can be regarded as
the
$L_\rs$-spin structure on the fibre bundle of Minkowski spaces 
\be
E_M=(LX\times M)/L\to\Si_{\rm T}
\ee
associated with the $L$-principal bundle
$LX\to\Si_{\rm T}$.
Since the principal bundles  $LX$ and $P_L$ (\ref{b3244}) are trivial, the
fibre bundle $E_M\to \Si_{\rm T}$ also is trivial, and this is
isomorphic to the pullback 
\beq
\Si_{\rm T}\op\times_X T^*X. \label{b3252}
\eeq

\begin{rem}
Since the bundle
$\Si_{\rm T}\to X$ is trivial, the fibre bundle $E_M$ is equivalent to the
trivial bundle of Minkoski spaces over the product
$S^3\times{\bf R}^7\times X$. It follows that the set of inequivalent spin
structures on the bundle $E_M$ is in bijective correspondence with the
cohomology group $H^1(S^3\times{\bf R}^7\times X;{\bf Z}_2)$ (\cite{law},
p.82). Since the cohomology group
$H^1(S^3;
{\bf Z}_2)$ is trivial and the spin structure on $S^3$ is unique \cite{dabr2},
one can show that inequivalent spin structures on
$E_M$ are classified by elements of the cohomology group $H^1(X;{\bf Z}^2)$
and, consequently, by inequivalent spin structures on $X$.
It follows that the spin structure (\ref{b3250}) on the fibre bundle
$E_M$ is unique. 
\end{rem}

Following the general discussion on spontaneous symmetry
breaking in the previous Section, let us consider the composite spinor bundle
\beq
S\ar^{\pi_{S\Si}}\Si_{\rm T}\ar^{\pi_{\Si X}} X, \label{L1}
\eeq
where 
\be
S=(\wt{LX}\times V)/L_\rs
\ee
is the spinor bundle $S\to \Si_{\rm T}$ associated with the
$L_\rs$-principal bundle $\wt{LX}\to \Si_{\rm T}$. 

Given a tetrad field $h$, there is the canonical isomorphism 
\be
i_h: S^h=(P^h\times V)/L_\rs \to (h^*\wt{LX}\times V)/L_\rs
\ee
of the $h$-associated spinor bundle $S^h$ (\ref{510}) to the
restriction $h^*S$ of the spinor bundle $S\to \Si_{\rm T}$ to $h(X)\subset
\Si_{\rm T}$. Then, every global section $s_h$ of the spinor bundle $S^h$
correspondes to the global section $i_h\circ s_h$ of the composite spinor 
bundle (\ref{L1}). Conversely, every global section $s$ of the composite
spinor bundle (\ref{L1}) which is projected onto a tetrad field
$h$ takes its values into the
subbundle $i_h(S^h)\subset S$.

Let the frame bundle $LX\to X$ be provided with a holonomic atlas 
(\ref{b3241}) and let the principal bundles $\wt{LX}\to \Si_{\rm T}$ and
$LX\to\Si_{\rm T}$ have the associated atlases $\{z^s_\e, U_\e\}$ and
$\{z_\e=\wt z\circ z^s_\e,U_\e\}$. With these atlases, the composite spinor
bundle
$S$ is equipped with the bundle coordinates $(x^\la,\si_a^\m, y^A)$, where
$(x^\la,
\si_a^\m)$ are coordinates of $\Si_{\rm T}$ such that
$\si^\m_a$ are the matrix components of the group element
$(T\f_\zeta\circ z_\e)(\si),$
$\si\in U_\e,\, \pi_{\Si X}(\si)\in U_\zeta.$
For each section $h$ of $\Si_{\rm T}$, we have
$(\si^\la_a\circ h)(x)= h^\la_a(x)$,
where $h^\la_a(x)$ are the tetrad functions (\ref{L6}).

The composite spinor bundle $S$ is equipped with the fibre spinor metric
\be
a_S(v,v')=\frac12(v^+\g^0v' +{v'}^+\g^0v), \qquad 
\pi_{S\Si}(v)=\pi_{S\Si}(v').
\ee

Since the fibre bundle of Minkowski spaces $E_M\to \Si_{\rm T}$ is isomorphic
to the pull-back bundle (\ref{b3252}), there exists the representation 
\beq
\bx{\g_\Si: T^*X\op\ot_{\Si_{\rm T}} S= (\wt{LX}\times (M\ot V))/L_\rs
\to (\wt{LX}\times\g(M\ot V))/L_\rs=S,} \label{L7}
\eeq
given by the coordinate expression
\be
\bx{\wh dx^\la=\g_\Si (dx^\la) =\si^\la_a\g^a.}
\ee
Restricted to $h(X)\subset \Si_{\rm T}$, this representation  recovers the
morphism
$\g_h$ (\ref{L4}).

Using this representation, one can construct the total Dirac
operator on the composite spinor bundle $S$ as follows. 

Since the composite fibre bundle (\ref{b3248}) is the composition of trivial
bundles 
\be
\wt{LX}:= L_\rs\times GL_4/{\rm L}\times X\to  GL_4/{\rm L}\times X\to X,
\ee
let us consider a principal connection $A_\Si$ (\ref{b3228}) on the
$L_\rs$-principal bundle $\wt{LX}\to
\Si_{\rm T}$ given
by the local connection form
\beq
A_\Si = (A_\la{}^{ab} dx^\la+ A^k_\m{}^{ab} d\si^\m_k)\ot L_{ab},
\label{L10}
\eeq
where
\ben
&& A_\la{}^{ab} =-\frac12 (\eta^{kb}\si^a_\m-\eta^{ka}\si^b_\m)
 \si^\nu_k K_\la{}^\m{}_\nu, \nonumber\\
&& A^k_\m{}^{ab}=\frac12(\eta^{kb}\si^a_\m -\eta^{ka}\si^b_\m) \label{M4}
\een
and $K$ is a world connection on $X$. We choose this connection 
because of the
following properties.

The principal connection (\ref{L10}) defines the associated spin connection
\beq
A_S = dx^\la\ot(\dr_\la + \frac12A_\la{}^{ab}L_{ab}{}^A{}_By^B\dr_A) +
d\si^\m_k\ot(\dr^k_\m +  \frac12A^k_\m{}^{ab}L_{ab}{}^A{}_By^B\dr_A)
\label{b3266}
\eeq
on the spinor bundle $S\to\Si_{\rm T}$. Let $h$ be a global section
of $\Si_{\rm T}\to X$ and $S^h$ the restriction of the bundle $S\to \Si_{\rm
T}$ to $h(X)$. It is readily observed that the restriction of the spin
connection (\ref{b3266}) to $S^h$ is exactly the spin connection
(\ref{b3216}).

The connection (\ref{b3266}) yields the first order differential
operator $\wt D$ (\ref{7.10}) on the composite spinor bundle $S\to X$. This
 reads
\ben 
&&\wt D:J^1S\to T^*X\op\ot_{\Si_{\rm T}} S,\nonumber\\
&&\wt D=dx^\la\ot[y^A_\la- \frac12(A_\la{}^{ab} + A^k_\m{}^{ab}\si_{\la
k}^\m)L_{ab}{}^A{}_By^B]\dr_A  =\label{7.10'} \\
&& \qquad dx^\la\ot[y^A_\la-
\frac14(\eta^{kb}\si^a_\m -\eta^{ka}\si^b_\m)(\si^\m_{\la k} -\si^\nu_k
K_\la{}^\m{}_\nu)L_{ab}{}^A{}_By^B]\dr_A. \nonumber
\een
 The corresponding restriction $\wt D_h$ (\ref{b3260}) of the
operator $\wt D$ (\ref{7.10'}) to
$J^1S^h\subset S^1S$ recovers the familiar covariant differential on the
$h$-associated spinor bundle $S^h\to X$ relative to the spin connection
(\ref{b3216}).

The composition of the representation (\ref{L7}) and the
differential (\ref{7.10'}) leads to the first order differential operator
\ben
&& \bx{\cD=\g_\Si\circ\wt D:J^1S\to T^*X\op\ot_{\Si_{\rm T}} S\to S,}
\label{b3261}\\
&& y^B\circ\cD=\si^\la_a\g^{aB}{}_A[y^A_\la-
\frac14(\eta^{kb}\si^a_\m -\eta^{ka}\si^b_\m)(\si^\m_{\la k} -\si^\nu_k
K_\la{}^\m{}_\nu)L_{ab}{}^A{}_By^B], \nonumber
\een
on the composite spinor bundle $S\to X$.
One can think of $\cD$ as being the  total Dirac operator  on
$S$ since, for every tetrad field $h$, the restriction of $\cD$ to
$J^1S^h\subset J^1S$  is exactly the Dirac operator $\cD_h$ (\ref{l13}) on
the
$h$-associated spinor bundle
$S^h$
in the presence of the background tetrad field $h$ and the spin connection
(\ref{b3216}).

It follows that gauge
gravitation theory is reduced to the model of metric-affine
gravity and Dirac fermion fields.

The total configuration space of this model is the jet manifold
$J^1Y$ of the bundle product
\beq
Y=(C_K\op\times_X\Si_{\rm T})\op\times_{\Si_{\rm T}} S=C_K
\op\times_{\Si_{\rm T}} S \label{042}
\eeq
coordinatized by $(x^\m,\si^\m_a, k_\m{}^\al{}_\bt,y^A)$, where 
$C_K$ is the bundle of world connections (\ref{015}).

Let $J^1_\Si Y$ denotes the first order jet manifold of the fibre bundle
$Y\to\Si_{\rm T}$. This fibre bundle can be provided with
the spin connection
\ben
&& A_Y: Y\ar J^1_\Si Y\ar^{\pr_2} J^1_\Si S, \nonumber\\
&&A_Y = dx^\la\ot(\dr_\la +\wt A_\la{}^{ab}L_{ab}^A{}_By^B\dr_A) +
d\si^\m_k\ot(\dr^k_\m +  A^k_\m{}^{ab}L_{ab}^A{}_By^B\dr_A), \label{b3263}
\een
where 
\be
\wt A_\la{}^{ab} =-\frac12 (\eta^{kb}\si^a_\m-\eta^{ka}\si^b_\m)
 \si^\nu_k k_\la{}^\m{}_\nu
\ee
and $A^k_\m{}^{ab}$ is given by the expression
(\ref{M4}). 

Using the connection (\ref{b3263}), we get the first order
differential operator
\ben 
&&\wt D_Y:J^1Y\to T^*X\op\ot_{\Si_{\rm T}} S,\nonumber\\
&&\wt D_Y=dx^\la\ot[y^A_\la- 
\frac14(\eta^{kb}\si^a_\m -\eta^{ka}\si^b_\m)(\si^\m_{\la k} -\si^\nu_k
k_\la{}^\m{}_\nu)L_{ab}{}^A{}_By^B]\dr_A \label{7.100}
\een
and the total Dirac operator
\ben
&& \cD_Y=\g_\Si\circ\wt D:J^1Y\to T^*X\op\ot_{\Si_{\rm T}} S\to S,
\label{b3264}\\
&& y^B\circ\cD=\si^\la_a\g^{aB}{}_A[y^A_\la-  \frac14(\eta^{kb}\si^a_\m
-\eta^{ka}\si^b_\m)(\si^\m_{\la k} -\si^\nu_k
k_\la{}^\m{}_\nu)L_{ab}{}^A{}_By^B]
\nonumber
\een
on  the fibre bundle $Y\to X$, where $\g_\Si$ denotes the
pull-back of the morphism (\ref{L7}) onto $Y\to (C_K\times\Si_{\rm T})$. 

Given a section
$K:X\to C_K$, the restrictions of the spin connection $A_Y$ (\ref{b3263}),
the operator $\wt D_Y$ (\ref{7.100}) and the Dirac operator $\cD_Y$
(\ref{b3264}) to $K^*Y$ are exactly the spin connection (\ref{b3266}) and the
operators (\ref{7.10'}) and (\ref{b3261}), respectively.

The total Lagrangian
density on the configuration space $J^1Y$ of the metric-affine gravity
and fermion fields is the sum
\beq
L=L_{\rm MA} + L_{\rm D} \label{060}
\eeq
of the metric-affine Lagrangian density 
\be
L_{\rm
MA}(k_\la{}^\al{}_\bt,\si^{\m\nu}), \qquad
\si^{\m\n}=\si^\m_a\si^\nu_b\eta^{ab},
\ee
and of the Dirac Lagrangian
density
\be
L_{\rm D}=[a_Y(i\cD_Y(w),w) -ma_S(w,w)]\si^0\w\cdots \w\si^3, 
\qquad w\in J^1S,
\ee
where $\si^a=\si^a_\m dx^\m$ and $a_Y$ is the pull-back of the fibre spinor
metric
$a_S$ onto the fibre bundle
$Y\to (C_K\times\Si_{\rm T})$.
Its coordinate expression is
\ben
&& \cL_{\rm D}=\{\frac{i}{2}\si^\la_q[y^+_A(\g^0\g^q)^A{}_B(y^B_\la-
\frac14(\eta^{kb}\si^a_\m
-\eta^{ka}\si^b_\m)(\si^\m_{\la k} -\si^\nu_k
k_\la{}^\m{}_\nu)L_{ab}{}^B{}_Cy^C)- \nonumber\\
&& \qquad (y^+_{\la A} -
\frac14(\eta^{kb}\si^a_\m
-\eta^{ka}\si^b_\m)(\si^\m_{\la k} -\si^\nu_k
k_\la{}^\m{}_\nu)y^+_C L^+_{ab}{}^C{}_A(\g^0\g^q)^A{}_By^B]- \label{b3265}\\ 
&&\qquad  my^+_A(\g^0)^A{}_By^B\}\sqrt{\nm\si}, \qquad \si=\det(\si_{\m\n}).
\nonumber
\een

It is readily observed that
\beq
\frac{\dr\cL_\psi}{\dr k^\m{}_{\nu\la}} + 
\frac{\dr\cL_\psi}{\dr k^\m{}_{\la\nu}} =0, \label{2C14}
\eeq
that is, the Dirac Lagrangian density (\ref{b3265}) depends on the torsion
\be
S_\m{}^\al{}_\nu= k_\m{}^\al{}_\nu -k_\nu{}^\al{}_\m
\ee
of a world connection.

\section{General covariant transformations}

Let us turn now to general covariant transformations.

Since the world manifold $X$ is parallelizable and the 
universal spin structure
is unique, the
$\wt{GL}_4$-principal bundle
$\wt{LX}\to X$, as like as the frame bundle $LX$, admits a canonical lift of
any diffeomorphism $f$ of the base $X$. This lift is 
defined by the commutative
diagram
\be
\begin{array}{rcccl}
 &\wt{LX} & \ar^{\wt \Phi} & \wt{LX}& \\
 _{\wt z} &\put(0,10){\vector(0,-1){20}} &  & \put(0,10){\vector(0,-1){20}} &
_{\wt z} \\
& LX & \ar^{\Phi} & LX & \\
 &\put(0,10){\vector(0,-1){20}} &  & \put(0,10){\vector(0,-1){20}} & \\
& X & \ar^f & X  &
\end{array} 
\ee
where $\Phi$ is the holonomic bundle automorphism of $LX$ (\ref{025}) induced
by $f$ \cite{dabr}.

The associated morphism of the spinor bunddle $S$ (\ref{L1}) is given by
the relation
\beq
\wt \Phi_S: [p, v]_{L_s} \to [\wt \Phi(p), v]_{L_s}, \qquad p\in
\wt{LX},\ v\in S. \label{b3270}
\eeq
Because $\wt \Phi$ is equivariant, this is a fibre-to-fibre automorphism
of the bundle $S\to \Si_{\rm T}$ over the canonical automorphism of the
$LX$-associated tetrad bundle $\Si_{\rm T}\to X$ (\ref{5.15}) which is
projected onto the diffeomorphism $f$ of $X$. Thus, we have the commutative
diagram of general covariant transformations of the spinor bundle $S$:
\be
\begin{array}{ccc}
 S & \ar^{\wt \Phi_S} & S \\
\put(0,10){\vector(0,-1){20}} &  & \put(0,10){\vector(0,-1){20}}  \\
\Si_{\rm T} & \ar^{\Phi_\Si} & \Si_{\rm T} \\
\put(0,10){\vector(0,-1){20}} &  & \put(0,10){\vector(0,-1){20}} \\
X & \ar^f & X  
\end{array} 
\ee

Accordingly, there exists a canonical lift $\wt\tau$ of every vector field
$\tau$ on
$X$ over $S$. The goal is to discover its coordinate expression. A
difficulty arises because tetrad coordinates $\si^\m_a$ 
of $\Si_{\rm T}$ depend
on the atlas of the bundle $LX\to \Si_{\rm T}$. Therefore,
non-canonical vertical components appear in the coordinate expression of
$\wt\tau$.

A comparison with the canonical lift (\ref{973}) 
of a vector field $\tau$ over
the metric bundle $\Si_{\rm PR}$ shows that the similar canonical lift of
$\tau$ over the tetrad bundle $\Si_{\rm T}$ 
coordinatized by $(x^\la,\si^\m_a)$
takes the form
\beq
\tau_\Si=\tau^\la\dr_\la + \dr_\nu\tau^\m \si^\nu_c \frac{\dr}{\dr \si^\m_c}
+Q^\m_c \frac{\dr}{\dr \si^\m_c}, \label{b3275}
\eeq
where the coefficients $Q^\m_c$ obey the conditions
\be
(Q^\m_a\si^\nu_b + Q^\nu_a\si^\m_b)\eta^{ab}= 0.
\ee
These coefficients
$Q^\m_a$ represent the above mentioned non-canonical part of the lift
(\ref{b3275}).

Let us consider a horizontal lift $\wt\tau_S$ of the vector field $\tau_\Si$
over the spinor bundle $S\to \Si_{\rm T}$ by means of the spin connection
(\ref{b3266}). Let $K$ be a symmetric connection whose 
geodesic vector field is $\tau$. We find that the canonical part of
the vector field
$\tau_\Si$ is lifted identically, and $\wt\tau_S$ reads
\be
&&\wt\tau_S = \tau^\la\dr_\la + \dr_\nu\tau^\m \si^\nu_c \frac{\dr}{\dr
\si^\m_c}+\\ 
&& \qquad Q^\m_c \frac{\dr}{\dr \si^\m_c} +\frac14  Q^\m_k(\eta^{kb}\si^a_\m
-\eta^{ka}\si^b_\m)(L_{ab}{}^A{}_By^B\dr_A + L^+_{ab}{}^A{}_By^+_A\dr^B).
\ee
This can be brought into the form
\be
&& \wt\tau_S = \tau^\la\dr_\la + 
\dr_\nu\tau^\m \si^\nu_c \frac{\dr}{\dr \si^\m_c} + \\
&& \qquad \frac14  Q^\m_k(\eta^{kb}\si^a_\m
-\eta^{ka}\si^b_\m)[-L_{ab}{}^d{}_c\si^\nu_d\frac{\dr}{\dr\si^\nu_c} +
L_{ab}{}^A{}_By^B\dr_A + L^+_{ab}{}^A{}_By^+_A\dr^B], 
\ee
where $L_{ab}{}^d{}_c$ are the generators (\ref{b3278}). The corresponding
total vector field on the bundle product $Y$ (\ref{042}) reads
\ben
&& \wt\tau_Y = \tau^\la\dr_\la + 
\dr_\nu\tau^\m \si^\nu_c \frac{\dr}{\dr \si^\m_c} + \label{b3279}\\
&&\qquad 
[\dr_\nu\tau^\al k_\m{}^\nu{}_\bt -
\dr_\bt\tau^\nu k_\m{}^\al{}_\nu - \dr_\m\tau^\nu
k_\nu{}^\al{}_\bt + \dr_{\m\bt}\tau^\al]\frac{\dr}{\dr k_\m{}^\al{}_\bt}+
\nonumber \\
&& \qquad \frac14  Q^\m_k(\eta^{kb}\si^a_\m
-\eta^{ka}\si^b_\m)[-L_{ab}{}^d{}_c\si^\nu_d\frac{\dr}{\dr\si^\nu_c} +
L_{ab}{}^A{}_By^B\dr_A + L^+_{ab}{}^A{}_By^+_A\dr^B]. \nonumber
\een

The Dirac Lagrangian density (\ref{060}), by construction, is invariant
separately under transformations of holonomic atlases of the frame
bundle $LX$ (passive covariant transformations acting on the Greek
indices) and under transformations of atlases of the principal bundles
$\wt{LX}\to\Si_{\rm T}$ and $LX\to\Si$ (passive spin and Lorentz gauge
transformations acting on the Latin indices). It follows that this Lagrangian
density is invariant under infinitesimal active gauge transformations 
whose generators are the vector fields (\ref{b3279}). Moreover, one can
exclude from calculations
 the last term in these vector fields which leads
to the N\"other flow and can, thus, consider only their 
canonical part 
\ben
&& \wt\tau = \tau^\la\dr_\la + 
\dr_\nu\tau^\m \si^\nu_c \frac{\dr}{\dr \si^\m_c} + \label{b3280}\\
&&\qquad 
[\dr_\nu\tau^\al k_\m{}^\nu{}_\bt -
\dr_\bt\tau^\nu k_\m{}^\al{}_\nu - \dr_\m\tau^\nu
k_\nu{}^\al{}_\bt + \dr_{\m\bt}\tau^\al]\frac{\dr}{\dr k_\m{}^\al{}_\bt}
\nonumber
\een
in order to obtain the energy-momentum conservation laws
\cite{giach95,giach97,sard97b}.


\begin{thebibliography}{ederf}


\bibitem{ar} A.Aringazin and A.Mikhailov, Matter fields in spacetime with
vector non-metricity, {\it Clas. Quant. Grav.}, {\bf 8}, 1685
(1991)

\bibitem{avis} S.Avis and C.Isham, Generalized spin structure on four
dimensional space-times, {\it Comm. Math. Phys.}, {\bf 72}, 103 (1980)

\bibitem{bor} A.Borowiec, M.Ferraris, M.Francaviglia and I.Volovich,
Energy-momentum complex for nonlinear gravitational Lagrangians in the
first-order formalism, {\it Gen. Rel. Grav.}, {\bf 26}, 637
(1994)

\bibitem{budi} P.Budinich and A.Trautman, {\it The Spinorial Chessboard}
(Springer-Verlag, Berlin, 1988)

\bibitem{corn} J.Cornwell, {\it Group Theory in Physics, Vol.II} (Academic
Press, London, 1984)

\bibitem{cra} J.Crawford, Clifford algebra: Notes on the spinor metric and
Lorentz, Poincar\'e and conformal groups, {\it J. Math.
Phys.}, {\bf 32}, 576 (1991)

\bibitem{dabr} L.Dabrowski and R.Percacci, Spinors and diffeomorphisms,
{\it Comm. Math. Phys.}, {\bf 106}, 691 (1986)

\bibitem{dabr2} L.Dabrowski and A.Trautman, Spinor structures on spheres and
projective spaces, {\it J. Math. Phys.}, {\bf 27}, 2022 (1986)

\bibitem{dods} C.Dodson, {\it Categories, Bundles and Spacetime Topology}
(Shiva Publishing Limited, Orpington, 1980)

\bibitem{fat95} F.Fatibene, M.Ferraris and M.Francaviglia and M.Godina, A
geometric definition of Lie derivative for spinor fields, in {\it
Differential Geometry and its Applications (Proceedings of the 6th
International Conference, Brno, August 28 -- September 1, 1995)}, eds 
J.Jany\v ska, I.Kol\'a\v r and J. Slov\v ak (Masaryk Univ., Brno, 1996), 549

\bibitem{fat95a} F.Fatibene, M.Ferraris and M.Francaviglia and M.Godina, Gauge
formalism for General Relativity and fermionic matter, {\it E-print
gr-qc/9609042}

\bibitem{ger} R.Geroch, Spinor structure of space-time in general relativity,
{\it J. Math. Phys.}, {\bf 9} 1739 (1968)

\bibitem{giach95} G.Giachetta and G.Sardanashvily, Stress-energy-momentum
tensors in Lagrangian field theory. 2. Gravitational superpotential, {\it
E-print: gr-qc/9511040}

\bibitem{giachcqg} G. Giachetta and G.Sardanashvily,
Stress-Energy-Momentum of Affine-Metric Gravity. Generalized Komar
Superportential, {\it Clas. Quant. Grav.}, {\bf 13} L67
(1996)

\bibitem{giach97} G.Giachetta and L.Mangiarotti, Dirac equation in
metric-affine gravitation theories and superpotentials, {\it
Int. J. Theor. Phys.} {\bf 36}, 125 (1997)

\bibitem{gor} F.Gordejuela and J.Masqu\'e, Gauge group and $G$-structures,
{\it J. Phys; A}, {\bf 28}, 497 (1995)

\bibitem{green} M.Greenberg, {\it Lectures on Algebraic Topology}
(W.A.Benjamin , Inc., Menlo Park, 1971)

\bibitem{greub} W.Greub and H.-R. Petry, On the lifting of structure groups,
in vol: {\it Differential Geometric Methods in Mathematical Physics II},
Lecture Notes in Mathematics (Springer-Verlag, Berlin, 1978), 217

\bibitem{guil} V.Guillemin and S.Stenberg, {\it Symplectic Techniques in
Physics} (Cambr. Univ. Press, Cambridge, 1990)

\bibitem{haw} S.Hawking and G.Ellis, {\it The Large Scale Structure of a
Space-Time} (Cambr. Univ. Press, Cambridge, 1973)

\bibitem{heh76} F.Hehl, P. von der Heyde, G.Kerlick and J.Nester, General
relativity with spin and torsion: Foundations and prospects, {\it Rev. Mod.
Phys.}, {\bf 48}, 393 (1976)

\bibitem{heh} F.Hehl, J.McCrea, E.Mielke and Y.Ne'eman: Metric-affine
gauge theory of gravity: field equations, Noether identities, world spinors,
and breaking of dilaton invariance, {\it Phys. Rep.} {\bf 258}, 1 (1995)

\bibitem{heu} B. van der Heuvel, Energy-momentum conservation in gauge
theories, {\it J. Math. Phys.}, {\bf 35}, 1668 (1994)

\bibitem{hir} F.Hirzebruch, {\it Topological Methods in Algebraic Geometry}
(Springer-Verlag, Berlin, 1966)

\bibitem{iva} D. Ivanenko and G. Sardanashvily,
The gauge treatment of gravity, {\it Phys. Rep.}, {\bf 94}, 1 (1983)

\bibitem{kob} S.Kobayashi and K.Nomizu, {\it Foundations of Differential
Geometry, Vol.1}  (John Wiley, N.Y. - Singapore, 1963).

\bibitem{kob72} S.Kobayashi, {\it Transformation Groups in Differential
Geometry} (Springer-Verlag, Berlin, 1972).
 
\bibitem{koss} Y.Kosmann, D\'eriv\'ees de Lie des spineurs, {\it Ann. di
Matematica Pura et Appl.}, {\bf 91}, 317 (1972)

\bibitem{law} H.Lawson and M-L. Michelson, {\it Spin Geometry} (Princeton
Univ. Press, Princeton, 1989)

\bibitem{marath} K.Marathe, A condition of paracompactness of a manifold,
{\it J. Diff. Geom.} {\bf 7}, 571 (1972)

\bibitem{nee} Y.Ne'eman and Dj.\v Sija\v cki, Unified affine gauge theory of
gravity and strong interactions with finite and infinite $\ol{GL}(4,{\bf R})$
spinor fields, {\it Annals of Physics}, {\bf 120}, 292 (1979)

\bibitem{nik} L.Nikolova and V.Rizov, Geometrical approach to the reduction of
gauge theories with spontaneous broken symmetries, {\it Rep. Math. Phys.},
{\bf 20}, 287 (1984)

\bibitem{nov} J.Novotn\'y, On the conservation laws in General Relativity, in
{\it Geometrical Methods in Physics. Proceeding of the Conference on
Differential Geometry and its Applications (Czechoslovakia 1983)}, ed. by
D.Krupka (University of J.E.Purkyn\v{e}, Brno, 1984), 207

\bibitem{nov93} J.Novotn\'y, Energy-momentum complex of gravitational field
in the Palatini formalism, {\it Int. J. Theor. Phys.}, {\bf 32}, 1033 (1993)

\bibitem{obu} Yu.Obukhov and S.Solodukhin, Dirac equation and the
Ivanenko--Landau--K\"ahler equation, {\it Int. J. Theor. Phys.}, {\bf 33},
225 (1994)

\bibitem{perc} R.Percacci, {\it Geometry on Nonlinear Field Theories}
(World Scientific, Singapore, 1986)

\bibitem{pon} V.Ponomarev and Yu.Obukhov, Generalized Einstein-Maxwell
theory, {\it Gen. Rel. Grav.}, {\bf 14}, 309 (1982)

\bibitem{rodr} W.Rodrigues and Q. de Souza, The Clifford bundle and the
nature of the gravitational field, {\it Found. Phys.}, {\bf 23}, 1465 (1993)

\bibitem{sard92} 
G.Sardanashvily, On the geometry of spontaneous symmetry breaking, {\it
J. Math. Phys.}, {\bf 33}, 1546 (1992)

\bibitem{sardz92}
G.Sardanashvily and O.Zakharov, {\it Gauge Gravitation Theory},
 (World Scientific, Singapore, 1992)

\bibitem{sardhp} G.Sardanashvily, {\it Gauge Theory in Jet Manifolds}
(Hadronic Press, Palm Harbor, 1993)

\bibitem{sard95}
G.Sardanashvily, {\it Generalized Hamiltonian Formalism for Field Theory.
Constraint Systems.} (World Scientific, Singapore, 1995)

\bibitem{sard97b}
G.Sardanashvily, Stress-energy-momentum conservation law in gauge gravitation
theory, {\it Clas. Quant. Grav.} {\bf 14}, 1371 (1997)

\bibitem{ste} N.Steenrod, {\it The Topology of Fibre Bundles} (Princeton Univ.
Press, Princeton, 1972)

\bibitem{swit} S.Switt, Natural bundles. II. Spin and the diffeomorphism
group, {\it J. Math. Phys.}, {\bf 34}, 3825 (1993)

\bibitem{tra} A.Trautman, {\it Differential Geometry for Physicists}
(Bibliopolis, Naples, 1984)

\bibitem{tuc} R.Tucker and C.Wang, Black holes with Weyl charge and
non-Riemannian waves, {\it Clas. Quant. Grav.}, {\bf 12}, 2587
(1995)

\bibitem{wist} G. Wiston, Topics on space-time topology, {\it Int. J. Theor.
Phys.}, {\bf 11}, 341 (1974); {\bf 12}, 225 (1975)

\bibitem{zul}  R.Zulanke and P.Wintgen, {\it Differentialgeometrie und
Faserb\"undel, Hochschulbucher fur Mathematik, Band 75}
(VEB Deutscher Verlag der Wissenschaften,  Berlin, 1972)

\end{thebibliography}
\end{document}